# First-principles Investigation of Exceptional Coarsening-resistant V-Sc(Al$_2$Cu)$_4$ Nanoprecipitates in Al-Cu-Mg-Ag-Sc Alloys

Junyuan Bai[1], Hao Xue[1], Jiaming Li[1], Xueyong Pang[1,3], Zhihao Zhao[1], Gang Liu[4*], Gaowu Qin[1,2*],

[1]*Key Laboratory for Anisotropy and Texture of Materials (Ministry of Education), School of Materials Science and Engineering, Northeastern University, Shenyang 110819, China*
[2]*Institute for Strategic Materials and Components, Shenyang University of Chemical Technology, Shenyang 110142, China*
[3]*Research Center for Metal Wires, Northeastern University, Shenyang 110819, China*
[4] *State Key Laboratory for Mechanical Behavior of Materials, Xi'an Jiaotong University, Xi'an, China.*

**Abstract:**

Aluminum-copper-magnesium-sliver (Al-Cu-Mg-Ag) alloys are extensively utilized in aerospace industries due to the formation of Ω nanoplates. However, the rapid coarsening of these nanoplates above 200 °C restricts their application at elevated temperatures. When introducing scandium (Sc) to these alloys, the service temperature of the resultant alloys can reach an unprecedented 400 °C, attributed to the in situ formation of a coarsening-resistant V-Sc(Al$_2$Cu)$_4$ phase within the Ω nanoplates. However, the fundamental thermodynamic properties and mechanisms behind the remarkable coarsening resistance of V nanoplates remain unexplored. Here, we employ first-principles calculations to investigate the phase stability of V-Sc(Al$_2$Cu)$_4$ phase, the basic kinetic features of V phase formation within Ω nanoplates, and the origins of the extremely high thermal stability of V nanoplates. Our results indicate that V-Sc(Al$_2$Cu)$_4$ is metastable and thermodynamically tends to evolve into a stable ScAl$_7$Cu$_5$ phase. We also demonstrate that kinetic factors are mainly responsible for the temperature dependence of V phase formation. Notably, the formation of V-Sc(Al$_2$Cu)$_4$ within Ω nanoplates modifies the Kagomé lattice in the shell layer of the Ω nanoplates, inhibiting further thickening of V nanoplates through the thickening pathway of Ω nanoplates. This interface transition leads to the exceptional coarsening resistance of the V nanoplates. Moreover, we also screened 14 promising element substitutions for Sc. These findings are anticipated to accelerate the development of high-performance Al alloys with superior heat resistance.

**Introduction**

As a preferred material for lightweight engineering, heat-treatable aluminum alloys are extensively utilized in transportation and aerospace due to their high specific strength and ease of processing for industrial production. By forming dense, coherent nanoprecipitates during aging treatments, such as the θ′-Al$_2$Cu phase [1–3] in Al-Cu-based alloys and η′/η$_2$ phases [4–6] in 7xxx series alloys, the mechanical properties of Al alloys can be significantly enhanced. Typically, these precipitation-strengthened Al alloys can satisfy performance requirements at room temperature. However, at elevated temperatures (often >200°C), the nanoprecipitates readily become unstable, which includes coarsening, dissolution, or replacement by other second phases, leading to a substantial decline in the mechanical performance of alloys [7–9]. For example, high-strength 7050 Al alloys retain only ~30% of their room temperature yield stress at 200°C due to the rapid coarsening of η$_2$-MgZn$_2$ nanoplates [10]. Similarly, while Ω nanoplates in heat-resistant Al-Cu-Mg-Ag alloys display excellent thermal stability below 200°C, their rapid coarsening beyond this temperature threshold also results in significant mechanical property degradation [11]. This poor thermal stability of nanoprecipitates limits the application of conventional high-strength heat-treatable Al alloys in high-temperature environments. Thus, currently, stabilizing these nanoprecipitates is a critical challenge for developing new creep-resistant Al alloys capable of operating at temperatures up to 300~400 °C.

Researchers have long pursued various strategies to stabilize coherent nanoprecipitates, expecting to inhibit their coarsening at elevated temperatures while maintaining a high volume fraction to ensure sufficient strength. Among these strategies, interfacial solute segregation has proven to be an effective approach, such as the coarsening of θ′-Al$_2$Cu nanoplates can be restricted by inhibiting interfacial diffusion through the segregation of solutes like Sc, Mn, or Zr to their broad terraces[7]. However, this approach is only viable below 350°C. To meet the demands of higher-temperature applications, Xue et al. [12] recently proposed a novel in-situ phase transformation stabilization strategy. Based on Al-Cu-Mg-Ag alloys, they found that the addition of Sc induces the in-situ formation of Sc(Al$_2$Cu)$_4$ phase (termed V phase, with a $D2_b$ Mn$_{12}$Th-prototype structure, space group: $I4/mmm$) within the {111}$_{Al}$-oriented Ω nanoplates, significantly enhancing the creep resistance of alloys and achieving an unprecedented tensile strength of ~100 MPa at 400°C.

According to our prior study [13], as indicated in Fig. 1, Ω nanoplate possesses a unique U$_T$|$K$|distorted θ-Al$_2$Cu|$K$|U$_T$ sandwich structure, with the outermost U$_T$|$K$ layers (where U$_T$ denotes the Undulating-Triangular lattice net and $K$ represents the Kagomé lattice net) serving exclusively as the shell and covering the core (i.e., distorted θ-Al$_2$Cu (designated θ$_d$-Al$_2$Cu) structure) part. A detailed description of these two lattice nets is provided in Fig. A1a. Xue et al.[12] reported that Sc atoms

diffuse into the Ω nanoplates via the ledges on the terrace of the nanoplates. Subsequently, the $\theta_d$-$Al_2Cu\rightarrow$V-Sc$(Al_2Cu)_4$ transition occurs upon the hopping of Sc atoms into the dodecahedral interstitial sites (marked by pink balls) within the $\theta_d$-$Al_2Cu$ structure. More importantly, when V nanoplates completely replace Ω nanoplates at temperatures above 400 °C, they exhibit exceptional thermal stability. Their size, number density, and volume fraction remain nearly unchanged after prolonged exposure at 400 °C. As a result, this in-situ phase transition stabilization strategy offers a promising avenue for developing future creep-resistant Al alloys.

Despite these advancements, three key issues need to be resolved before this in-situ phase transition stabilization strategy becomes a universal one: (1) The formation of the V-Sc$(Al_2Cu)_4$ phase within Ω nanoplates requires temperatures exceeding 300 °C, which is remarkably higher than the typical formation temperatures of other nanoprecipitates like $\eta$-$MgZn_2$ [4], Ω[14], and $T_1$ [15] (often ~200 °C). It is yet to be determined whether the temperature-dependent formation of V phase is governed by the thermodynamics, kinetics, or a synergistic interaction of both factors. (2) The V-Sc$(Al_2Cu)_4$ nanoplates exhibit extraordinary thermal stability at elevated temperatures, with almost no coarsening phenomenon observed. Nonetheless, the atomistic mechanisms behind this stability are still poorly understood. (3) The introduction of Sc alters the thermodynamics of the Al-Cu-Mg-Ag alloy, inducing the V-Sc$(Al_2Cu)_4$ phase formation. However, the high cost of Sc hinders the widespread application of this alloy. Therefore, it is crucial to explore if more economical elements, such as different cheap rare earth (RE) elements, can serve as viable alternatives to Sc.

The present study addresses these issues through the use of accurate first-principles calculations. This computational method has been extensively employed to investigate the thermodynamic and kinetic properties of precipitate phases in Al alloys [2,16–18] and other systems [13,19–21]. By constructing 0 K and high-temperature Al-Cu-Sc convex hull diagrams, we evaluated the phase stability of the V-Sc$(Al_2Cu)_4$ phase, finding it to be unexpectedly metastable. Furthermore, using a cluster expansion method, we searched for a new ternary stable phase, $ScAl_7Cu_5$, at the Al-rich corner, which is the evolution destination of the metastable V-Sc$(Al_2Cu)_4$ phase. Through evaluating the activation energy for Sc hopping into dodecahedral interstitial positions, the basic kinetic features of V phase formation within Ω nanoplates were elucidated. Notably, we unraveled the origin of the high coarsening resistance in V nanoplates by determining the complete atomic configurations of sandwich V nanoplates, and the potential major alloying elements that could substitute for Sc were also screened. These findings aim to deepen the understanding of the in-situ phase transition stabilization strategy and advance the development of high-performance Al alloys with superior creep resistance.

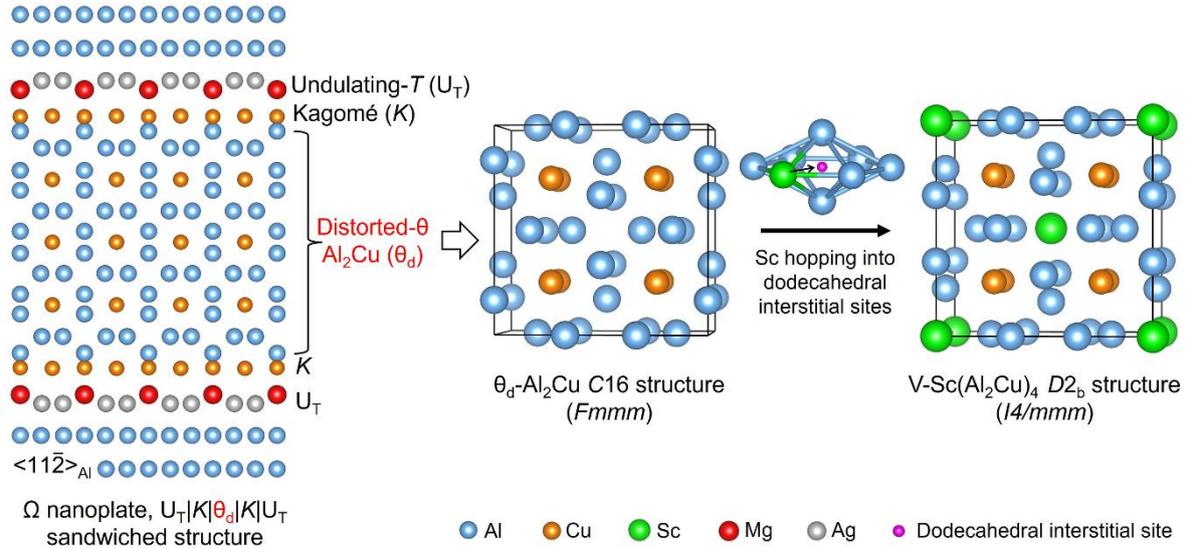

**Figure 1.** Atomic structures of Ω nanoplates, the $θ_d$-$Al_2Cu$ phase, and the V-$Sc(Al_2Cu)_4$ phases, along with schematic illustrations of the $θ_d$-$Al_2Cu$→V-$Sc(Al_2Cu)_4$ transition, facilitated by Sc hopping into dodecahedral interstitial sites within the $θ_d$-$Al_2Cu$ $C$16 structure.

## 2. Methods

2.1. DFT calculations

First-principles calculations based on density functional theory (DFT) were performed using the Vienna *ab initio* simulation package (VASP)[22,23] with Blochl's projector augmented wave (PAW) potential method [24]. The exchange-correlation energy functional was described with the generalized gradient approximation (GGA) as parameterized by Perdew-Burke-Ernzerhof (PBE) [25]. A 520 eV plane wave cutoff was adopted with the convergence criteria for energy and the atomic force set as $10^{-6}$ eV and $10^{-2}$ eV/Å, respectively. Partial occupancies were determined by using the first-order Methfessel-Paxton method with a smearing width of 0.2 eV [26]. A Γ-centered $k$-mesh with a spacing of 0.25 Å$^{-1}$ between $k$ points was employed, ensuring energy convergence to better than 1 meV/atom. Structures were visualized using the VESTA software [27].

In the calculation of formation energies for various antisite defects in the $Al_2Cu$ (θ), $ScAl_3$, $Sc(Al_2Cu)_4$, and $ScAl_7Cu_5$ phases, we employed a 96-atom model (2×2×2 $Al_2Cu$-θ unit-cell) for the $Al_2Cu$ (θ) phase, a 36-atom model (2×2×2 $ScAl_3$ unit-cell) for the $ScAl_3$ phase, and a 208-atom model (2×2×2 $Sc(Al_2Cu)_4$ and $ScAl_7Cu_5$ unit-cell) for the $Sc(Al_2Cu)_4$ and $ScAl_7Cu_5$ phases. In the calculation of activation energies ($E_a$) for Sc hopping into dodecahedral interstitial sites, we employed a 432-atom model (3×3×2 $Al_2Cu$-$θ_d$ unit-cell) and a 396-atom model (2×4×2 $Al_2Cu$-$θ_d$ unit-cell) to evaluate Sc hopping within the Ω nanoplates during the incubation and late-growth stages, respectively. This type of model was chosen because the Al|Ω|Al sandwich models (see Fig. S1b), while more realistic, would introduce a vacuum layer in the model that inevitably causes discontinuities in the electronic density and significantly slows the convergence of electronic self-consistency. Comparison of $E_a$ values for Sc substitutional diffusion in the 3×3×2 model and the Al|Ω|Al sandwich model (see Fig. S1) showed that the former yields $E_a$ values of 0.682 eV, while the latter computes 0.689 eV, with only 0.007 eV difference. This negligible discrepancy validates the accuracy of our chosen models in replicating $E_a$ values of the latter mode, thus confirming their reliability. Furthermore, all results obtained in this study have been rigorously tested to ensure the exclusion of any influence from model size.

2.2. Concentrations and formation energies of antisite defects

In the dilute limit, the concentration of an antisite defect is determined by the formation energy $\Delta G_f$ through a Boltzmann expression [28]:

$$c = N_{site} \exp\left(\frac{-\Delta G_f}{k_B T}\right) \tag{1}$$

$N_{site}$ is the number of sites in the crystal where the defect can occur (per unit volume), $k_B$ is the Boltzmann constant, $T$ is the temperature, and $\Delta G_f$ corresponds to

$$\Delta G_f = \Delta E_f - T\Delta S_f + P\Delta V_f \tag{2}$$

Here, $\Delta E_f$ is the change in total energy, $\Delta S_f$ is mainly the change in vibrational entropy, and $\Delta V_f$ is the change in volume when the antisite defect is introduced into the system. Since the contribution of volume changes is relatively small and the changes in entropy are of the same order when comparing different defects, we focus only on computing formation energy terms. The formation energy of an antisite defect, $\Delta E_f$, is computed as [29]:

$$\Delta E_f = E_{tot}^{antisite} - E_{tot}^{bulk} - \sum_{i=1}^{n} n_i \mu_i \tag{3}$$

$E_{tot}^{antisite}$ is the total energy derived from a supercell calculation containing an antisite defect, $E_{tot}^{bulk}$ is the total energy for the perfect crystal using an equivalent supercell, the integer $n_i$ indicates the number of atoms of host atoms or defect atoms that have been added to ($n_i > 0$) or removed from ($n_i < 0$) the supercell to form the defect, and $\mu_i$ are the chemical potentials representing the energy of the reservoirs with which atoms are being replaced. The $\mu_i$ of Al, Cu, and Sc in the $Al_2Cu$-θ, $ScAl_3$, $Sc(Al_2Cu)_4$, and $ScAl_7Cu_5$ phases were determined by solving the following equations based on equilibrium relations between various phases reported in experiments [12,30]:

$$\mu_{Al} = E_{FCC-Al}, \quad 2\mu_{Al} + \mu_{Cu} = E_{Al_2Cu(\theta)};$$

$$3\mu_{Al} + \mu_{Sc}^{ScAl_3} = E_{ScAl_3};$$

$$\mu_{Sc}^{Sc(Al_2Cu)_4} + 8\mu_{Al} + 4\mu_{Cu} = E_{Sc(Al_2Cu)_4};$$

$$\mu_{Sc}^{ScAl_7Cu_5} + 7\mu_{Al} + 5\mu_{Cu} = E_{ScAl_7Cu_5} \tag{4}$$

The calculated formation energies of various antisite defects in the $Sc(Al_2Cu)_4$ and $ScAl_7Cu_5$ phases are presented in Fig. 3, whereas the formation energies of antisite defects in $Al_2Cu-\theta$ and $ScAl_3$ are shown in Fig. S4 and Table S1.

2.3. Calculations of vibrational and configurational entropies

In thermodynamic equilibrium, defects with a positive formation energy $\Delta E_f$ are stabilized by their configurational entropy

$$S_{conf} = k_B \ln W \tag{5}$$

where the number of microstates $W$ for $n_i$ antisite defects of a specific type on $N$ lattice sites is given by

$$W = \prod_i W_i = \frac{N!}{(N - \sum_{i=1}^{n} n_i)! \prod_{i=1}^{n} n_i!} \tag{6}$$

The consideration of Eqs. (5) and (6) in the thermodynamic limit allow the application of the Stirling approximation and normalization of $S^{conf}$ to a per atom by dividing $N$, finally resulting in

$$\frac{S_{conf}}{N} = -k_B \sum_{i=1}^{n} c_i \ln c_i \tag{7}$$

Where $c_i = n_i / N_i$.

Typically, the entropy associated with harmonic atomic vibrations at high temperatures is proportional to the logarithmic moment of the phonon densities of states (DOS), $g(\omega)$:

$$S_{vib} = k_B \int_0^\infty \left[ 1 + \ln\left(\frac{k_B T}{\hbar \omega}\right) + \ldots \right] g(\omega) d\omega \tag{8}$$

Where $\hbar$ is the reduced Planck's constant, and $\omega$ is the phonon frequencies. Higher-order terms in Eq. (8) are negligibly small for temperatures higher than the characteristic Debye temperature $\theta_D$ of atomic vibrations. Because typical metals $\theta_D$ are well below room temperature, Eq. (8) can be truncated after the first two terms. Thus, the contribution of $S_{vib}$ and $S_{conf}$ to the free energy can be obtained by $E_{vib+conf} = -T(S_{vib} + S_{conf})$. For phonon dispersion and $g(\omega)$ calculations, the harmonic interatomic force constants were obtained by density functional perturbation theory (DFPT) [31] using the supercell approach, which calculates the dynamical matrix through the linear response of electron density. Then, phonon dispersion and $g(\omega)$ were computed using the *Phonopy* code [32] with the obtained harmonic interatomic force constants as input.

2.4. Cluster-expansion approach searching for stable structures

We utilized the data from the Open Quantum Material Database (OQMD)[33] and the Materials Project[34] databases to reconstruct the ternary convex hull for the Al-Cu-Sc system. All phases involved in this ternary system were recalculated in this study. By employing the *pymatgen* code[35], we utilized the recalculated DFT energies to plot the ternary Al-Cu-Sc convex hulls at 0 K and elevated temperatures to assess phase stability. It's important to note that the zero-temperature convex hull considers only internal energy, while high-temperature convex hulls also incorporate vibrational and/or configurational entropy contributions. To determine the atomic structure of the stable $\tau$-ScAl$_{8-x}$Cu$_x$ phase in the Al-rich corner, we employed a cluster expansion (CE) approach. In this CE calculation, 60 effective cluster interaction (ECI) coefficients were used to achieve the final cluster expansion for $\tau$-ScAl$_{8-x}$Cu$_x$ by fitting 235 ordered input structures (all input structures containing 26 atoms per cell). The cluster expansion has a cross-validation score of 0.008 eV/mixing atom. The CE method was implemented using MAPS code in the Alloy Theoretical Automated Toolkit (ATAT) software [36].

2.4. Evaluations of hopping incubation time for interstitial diffusion events

To determine the position of the saddle point (i.e., activation energy $E_a$) and the associated minimum energy pathway during atomic hopping, calculations were performed employing the climbing image nudged elastic band (CI-NEB) method [37,38] with two interpolated images. According to Vineyard's harmonic transition state theory[39], the hopping incubation time for a diffusion event can be written as: $t = v_0^{-1} \exp[\frac{\Delta E_a}{k_B T}]$, where $k_B$ denotes the Boltzmann constant, $\Delta E_a$ represents the activation energy, and $v_0$ is the attempt frequency. Here, the hopping incubation time $t$ was evaluated at T=200~400 °C. For a system comprising $N$ atoms, $v_0$ is given by: $v_0 = \prod_{l=1}^{3N-3} v_l / \prod_{k=1}^{3N-4} \tilde{v}_k$, where the frequencies $v_l$ signifies the 3$N$-3 normal-mode vibrational frequencies of the initial configuration, and $\tilde{v}_k$ are the 3$N$-4 non-imaginary normal-mode vibrational frequencies at the saddle-point configuration. Notably, it is currently infeasible to compute the exact $E_a$ for every possible hopping across all local atomic configurations during the late-growth stages. Therefore, the results for the late-growth stages can only be used for qualitative comparison with those of the incubation stage.

2.5. *Ab initio* molecular dynamics (AIMD) simulations

Born-Oppenheimer molecular dynamics are used for the AIMD simulations such that the electronic and ionic subsystems are fully decoupled. The energy tolerance for the electronic relaxation is set as 1×10$^{-5}$ eV. All the simulations were performed in a canonical NVT ensemble (i.e., keeping constant atom number, volume, and temperature) with temperature (500 K) controlled by the Nosé thermostat[40,41]. The total number of time steps is 5000, with each time step corresponding to 2 fs.

2.6. Simulation of HAADF-STEM images

The high-angle annular dark-field scanning transmission electron microscopy (HAADF-STEM) images were simulated with the code of QSTEM [42], and the corresponding parameters used for simulations are: a spherical aberration coefficient $C_3$= 1 mm, defocus values $\Delta f$ = = -60 nm beam convergence angle $\alpha$= 15 mrad, annular dark-field detector range is from 70 to 200 mrad.

## 3. Results and Discussions:

### 3.1 Phase stability of the V-Sc(Al$_2$Cu)$_4$ phase at 0 K and high temperatures

The formation of the V-Sc(Al$_2$Cu)$_4$ phase within Ω nanoplates occurs only at temperatures beyond 300°C. To elucidate the key factors contributing to this temperature-dependent formation, we first examine thermodynamic factors by assessing the phase stability of the V-Sc(Al$_2$Cu)$_4$ phase through the construction of Al-Cu-Sc convex hull diagrams at different temperatures. Convex hull analysis is a powerful and widely utilized methodology for assessing phase stability[43]. A phase is considered thermodynamically stable (or a "ground state") if its formation energy at a given composition is lower than that of all other phases with the same composition, as well as any linear combination of energies of other phases in the phase space. The collection of all ground states in a given phase space constitutes the convex hull, while phases located above the convex hull are thermodynamically unstable. According to convex hull theory, in binary or ternary systems, unstable phases whose compositions lie between stable phases will decompose into their neighboring stable phases, while those above a stable phase with identical composition will transform into that stable phase.

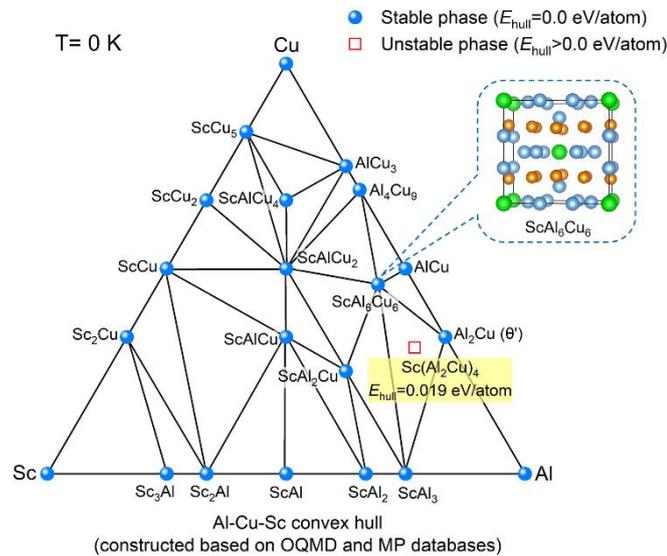

**Figure 2**. The 0 K convex hull of the Al-Cu-Sc system was constructed using data sourced from the OQMD [33]and Materials Project (MP)[34] databases. An inset within the figure presents the ScAl$_6$Cu$_6$ structure, where Al, Cu, and Sc atoms are symbolized by light-blue, yellow, and green spheres, respectively.

Utilizing data from the OQMD [33] and Materials Project (MP) [34] databases, we reconstructed the zero-temperature Al-Cu-Sc convex hull (Fig. 2), marking the stable phases with light-blue spheres. Our results indicate that the V-Sc(Al$_2$Cu)$_4$ phase is unstable (or metastable) at 0 K, with a hull distance ($E_{hull}$) of 0.019 eV/atom. Additionally, the V-Sc(Al$_2$Cu)$_4$ phase falls outside the Al-ScAl$_3$-Al$_2$Cu (θ′) triangle, indicating it cannot precipitate from common dilute alloys upon heating. At the same time, although the stable $ScAl_6Cu_6$ phase in this ternary system has a $D2_b$ structure similar to that of the V-$Sc(Al_2Cu)_4$ phase, it does not form a tie-line with Al. This means that this phase also cannot achieve equilibrium with the matrix Al and thus cannot precipitate from the supersaturated Al matrix during heating. Due to the 0 K ternary convex hull diagram considering only the internal energy contribution and neglecting the entropic effects, it is not difficult to understand these results that contradict experimental observations.

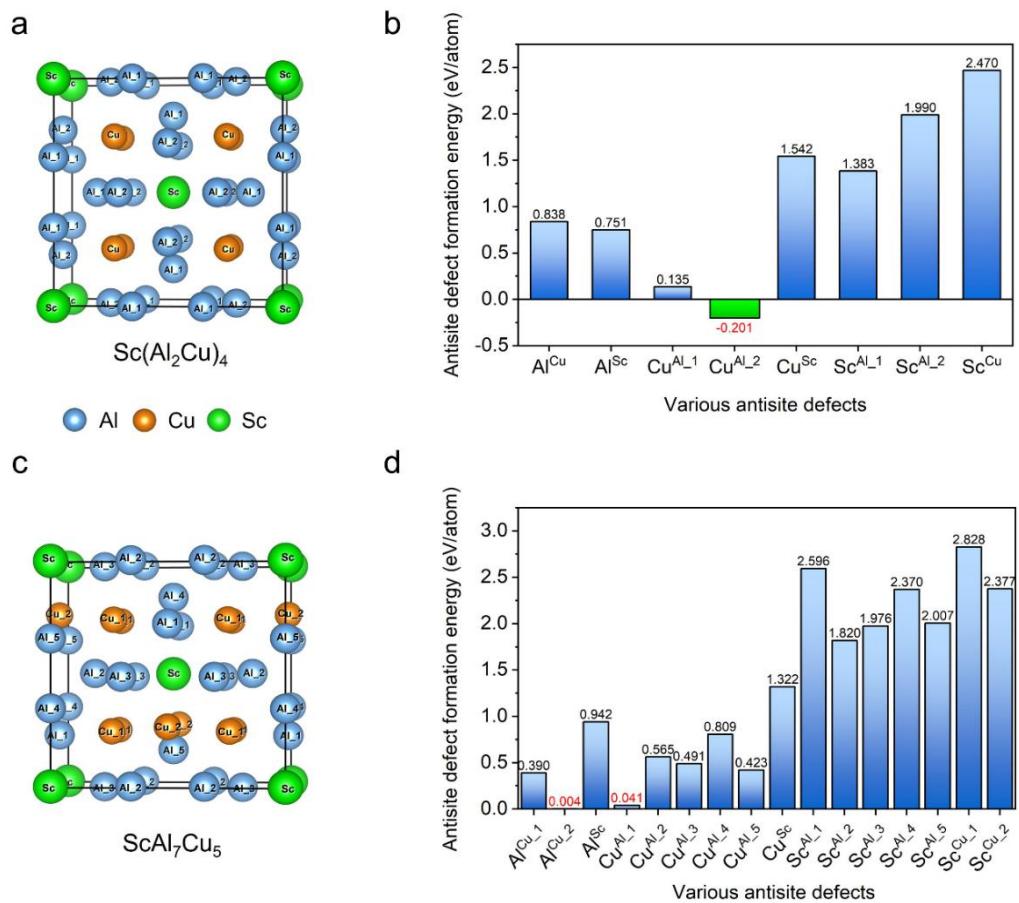

**Figure 3.** Formation energies of various antisite defects in the $Sc(Al_2Cu)_4$ (a-b) and $ScAl_7Cu_5$ (c-d) structures. The different symmetrical atoms in the $Sc(Al_2Cu)_4$ and $ScAl_7Cu_5$ are classified and annotated on the respective atomic sites.

At elevated temperatures, entropic contributions, particularly vibrational entropy ($S_{vib}$), may significantly influence phase stability, causing unstable phases at low temperatures to become stable at elevated temperatures[44]. A well-known example is the vibrational entropic stabilization of the θ-$Al_2Cu$ phase [45] relative to its metastable counterpart, θ′-$Al_2Cu$. At 0 K, the θ′-$Al_2Cu$ phase is stable because of its lower formation energy. However, as temperatures above ~ 200 °C, θ-$Al_2Cu$ becomes stable due to its larger $S_{vib}$ contributions. While electronic entropy ($S_{ele}$) and configurational entropy

($S_{conf}$) also contribute to phase stability, $S_{ele}$ is often neglected due to its minor contributions, and $S_{conf}$ is typically considered in multicomponent systems [46] such as high-entropy alloys. As a result, $S_{vib}$ is often the primary focus when assessing phase stability at elevated temperatures, with less attention paid to $S_{conf}$, especially for complex-structured intermetallic compounds [44]. Nevertheless, this study reveals that $S_{conf}$ can play a major role in enhancing phase stability when the formation energy of certain antisite defects in the corresponding phase is sufficiently low to dramatically increase $S_{conf}$.

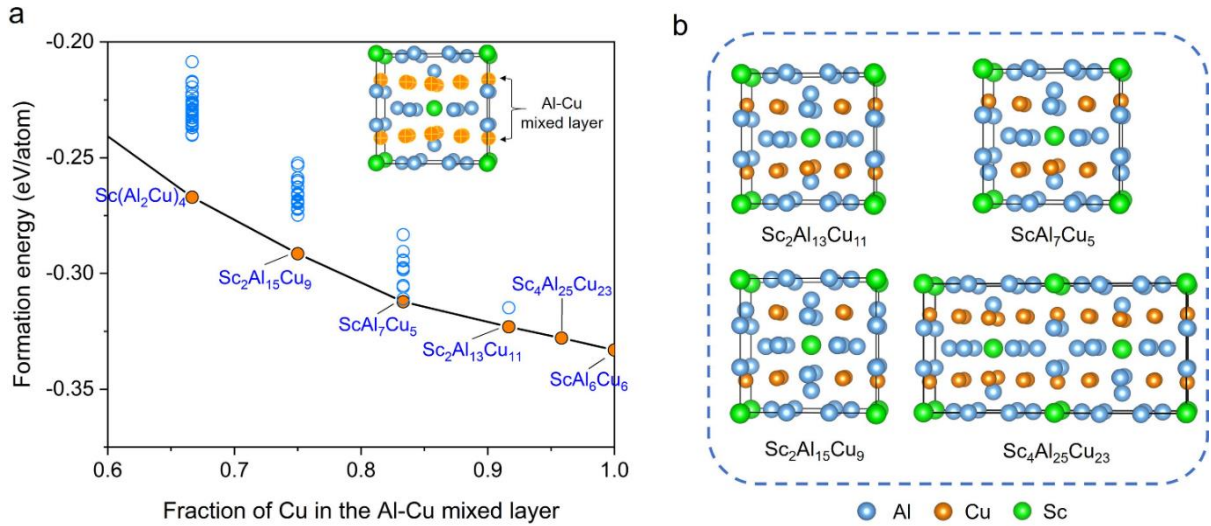

**Figure 4**. a. The variation of formation energies of generated structures as a function of Cu fraction in the Al-Cu mixed layer. b. Atomic configurations of the $Sc_2Al_{13}Cu_{11}$, $ScAl_7Cu_5$, $Sc_2Al_{15}Cu_9$, and $Sc_4Al_{25}Cu_{23}$ phases.

For the V-$Sc(Al_2Cu)_4$ phase, we found that it remains metastable at elevated temperatures due to the following reasons. First, our established Al-rich corner Al-Cu-Sc convex hull at 500~800 K (see Fig. A2), which accounts only for $S_{vib}$ (detailed calculation procedures in Section 2.3), shows little deviation from the 0 K convex hull, with V-$Sc(Al_2Cu)_4$ still keeping unstable and its position falling within the $ScAl_6Cu_6$-$ScAl_3$-$Al_2Cu$ triangle. This indicates that $S_{vib}$ alone cannot provide sufficient driving force for V-$Sc(Al_2Cu)_4$ (i.e., substantially lower its formation energy) to transition into a stable phase and form a tie-line with Al. More crucially, when we further include $S_{conf}$ by calculating the formation energy of various antisite defects (more details in Section 2.2), we unexpectedly discovered that the formation energy of the $Cu^{Al\_2}$ defect (where a Cu atom occupies an Al_2 site in the $Sc(Al_2Cu)_4$ structure) is negative, as shown in Fig. 3a-b. Typically, antisite defect formation energies for stable phases are positive[47]. The negative value for the $Cu^{Al\_2}$ defect indicates that the $Sc(Al_2Cu)_4$ structure is inherently unstable, with Al_2 sites tending to be occupied by Cu. As a result, the V-$Sc(Al_2Cu)_4$ phase remains metastable at elevated temperatures and is expected to evolve into a specific stable structure.

In experimental Al-rich corner Al-Cu-Sc phase diagrams at 400~500°C [48,49], a stable ternary $Al_{8-x}Cu_{4+x}Sc$ (0<x<2.6) compound (commonly referred to as the τ phase) has been explicitly documented. This τ phase can coexist with the Al solid solution (designated (Al)), forming a τ-(Al) two-phase region in the phase diagram. Hence, the experimentally identified τ phase should not only lie on the convex hull but also form a tie-line with Al. However, the exact stoichiometry of the τ-$Al_{8-x}Cu_{4+x}Sc$ phase remains uncertain experimentally. To identify the structure of the stable τ phase, which represents the evolution endpoint of the metastable V-$Sc(Al_2Cu)_4$ phase, we employed a cluster expansion approach. Given that Cu atoms energetically prefer to occupy the Al_2 sites within the V-$Sc(Al_2Cu)_4$ structure (see Fig. 3a), we varied Cu compositions at both the Al_2 and Cu sites (designated as the Al-Cu mixed layer in the inset of Fig. 4a) to search for the stable τ structure. Figure 4a presents the variation in formation energies of generated structures as a function of Cu fraction in the Al-Cu mixed layer, with the six structures displaying the lowest formation energy for each composition highlighted with orange circles. Among these, $Sc(Al_2Cu)_4$ and $ScAl_6Cu_6$ are known structures, while the remaining $Sc_2Al_{15}Cu_9$, $ScAl_7Cu_5$, $Sc_2Al_{13}Cu_{11}$, and $Sc_4Al_{25}Cu_{23}$ (see Fig. 4b) are newly screened.

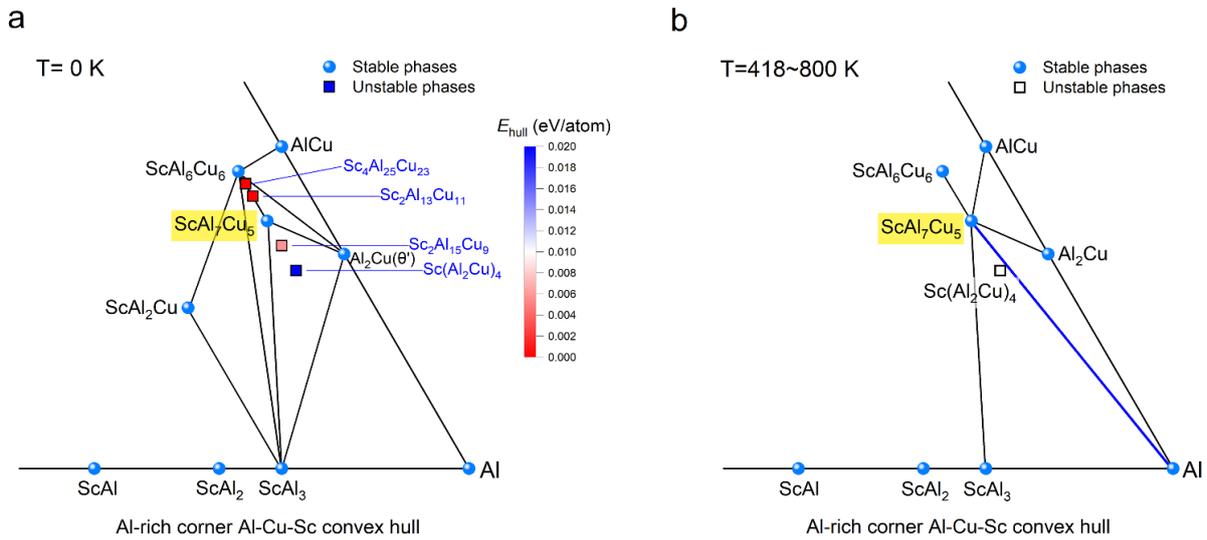

**Figure 5.** The convex hull for the Al-Cu-Sc system at T= 0 K (a) and T=418~800 K (b).

The Al-rich corner Al-Cu-Sc convex hull at 0 K (Fig. 5a), including the four newly screened structures, reveals that the $ScAl_7Cu_5$ is the stable phase and does not form a tie-line with Al. The remaining three phases, although unstable, exhibit relatively low $E_{hull}$ values. Similarly, when reconstructing the Al-rich corner Al-Cu-Sc convex hull at elevated temperatures, incorporating only the $S_{vib}$ is also insufficient to enable the formation of a tie-line between the stable $ScAl_7Cu_5$ phase and Al (see Fig. A3). This indicates that $S_{conf}$ needs to be taken into account. Specifically, during calculating $S_{conf}$, despite positive formation energies for various antisite defects in the $ScAl_7Cu_5$ structure (see Fig. 3c-d), the $Al^{Cu\_2}$ (0.004 eV/atom) and $Cu^{Al\_1}$ (0.041 eV/atom) defects exhibit significantly lower values compared to other kinds of antisite defects and those in the $Al_2Cu$ (θ) and $ScAl_3$ phases (see Table. S1). According to Eq. (1), such low antisite defect formation energies can result in high defect concentrations, yielding a non-negligible $S_{conf}$. Consequently, only the $S_{conf}$ of the ScAl$_7$Cu$_5$ phase was considered, with the corresponding Al-rich corner Al-Cu-Sc convex hull at elevated temperatures presented in Fig. 5b.

Our results indicate that the incorporation of $S_{conf}$ indeed provides a sufficient driving force for the stable ScAl$_7$Cu$_5$ phase to form a tie-line (highlighted by blue) with Al at temperatures between 418~800 K. Hence, we suggest that the stable $\tau$ phase is of ScAl$_7$Cu$_5$ structure, and the observed variable stoichiometry in $\tau$-Al$_{8-x}$Cu$_{4+x}$Sc compounds primarily arises from their high concentrations of Al$^{Cu\_2}$ and Cu$^{Al\_1}$ antisite defects. Additionally, the metastable V-Sc(Al$_2$Cu)$_4$, which is kinetically favored due to the θ$_d$-Al$_2$Cu structure in the Ω nanoplates core, thermodynamically tends to evolve into the stable ScAl$_7$Cu$_5$ structure with prolonged heating. However, our simulated HAADF-STEM images (Fig. S5) of sandwich nanoplates containing ScAl$_7$Cu$_5$ and ScAl$_6$Cu$_6$ phases display little difference from experimental observations. This means that the current HAADF-STEM observations are insufficient to clearly differentiate variations in Cu compositions in the Al-Cu mixed layers and thus cannot accurately determine the extent of compositional evolution in the V phase. At the same time, even for the stable ScAl$_7$Cu$_5$ phase, a tie-line with Al can form at temperatures just above 418 K (>145 °C), without requiring much higher temperatures. Therefore, thermodynamics is not a limiting factor in the temperature-dependent formation of V-Sc(Al$_2$Cu)$_4$/$\tau$-ScAl$_7$Cu$_5$ phases.

**3.2 Kinetics lead to temperature dependence of V-Sc(Al$_2$Cu)$_4$ formation**

In their experimental study, Xue et al. [12] observed that Sc incorporation into Ω nanoplates occurs predominantly via adsorption of Sc atoms at ledges on coherent terraces. This phenomenon can be attributed to the strain fields generated by partial dislocations at ledge edges, which facilitate Sc diffusion through interfacial defects into the Ω nanoplates. Upon entering the θ$_d$-Al$_2$Cu core of Ω nanoplates, the V-Sc(Al$_2$Cu)$_4$ structure forms as Sc hops into the dodecahedral interstitial sites at a specific moment. This mechanism challenges conventional recognition that interstitial diffusion in metals is typically limited to smaller atoms like carbon, hydrogen, and oxygen [50–52]. Owing to the high activation energy ($E_a$) required, large-sized Sc atoms are expected to display this interstitial diffusion rarely. However, the nucleation of a new phase is often accomplished by such low-probability, rare events [53–55]. By evaluating the hopping incubation time (see Fig. 6) of Sc into the dodecahedral interstitial sites during the incubation and late-growth stages of V-Sc(Al$_2$Cu)$_4$ formation (more computational details in Section 2.4), we show that Sc interstitial diffusion is a rare event and a major factor causing temperature-dependent formation of the V phase.

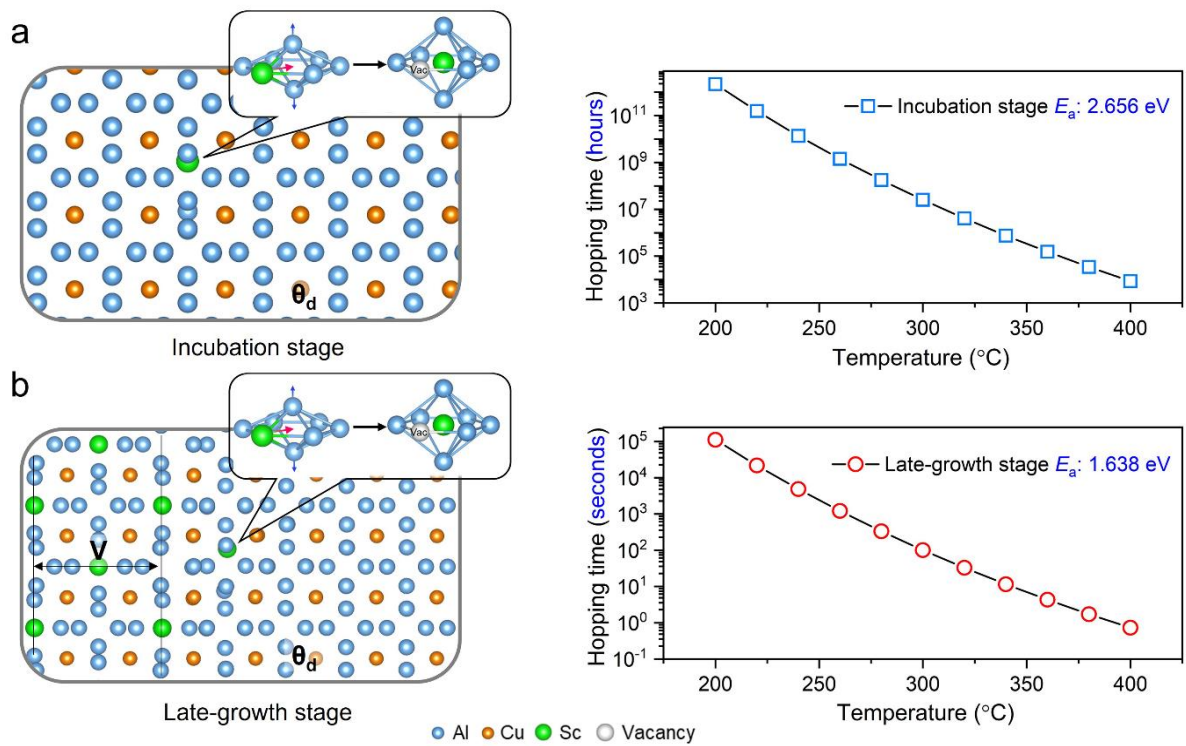

**Figure 6**. The hopping incubation time of the Sc atom during the incubation (a) and late-growth (b) stages of the V-Sc(Al$_2$Cu)$_4$ formation as a function of temperature. The left panel in the figure presents detailed scenarios depicting these two different stages.

In Supplementary Text 1, we demonstrate that the major interstitial hopping mode for Sc, as depicted in the inset of the left panel of Fig. 6, involves Sc hopping to dodecahedral interstitial positions, leaving a vacancy at its original site. The detailed scenarios for incubation and late-growth stages are shown in the left panel of Fig. 6, while the right panel presents the hopping incubation time of the Sc atom as a function of temperature (ranging from 200~400°C), with incubation and late-growth stages measured in hours and seconds, respectively. During the incubation stage (see Fig. 6a), as temperature increases from 200 °C to 400 °C, the hopping time reduces from an initial $10^{12}$ magnitudes to several thousand hours. This indicates that the first interstitial hopping event of Sc within the θ$_d$ structure demands an extremely long incubation time, even at 400 °C. In the late-growth stage (see Fig. 6b), due to the distortion induced by the formation of the V-Sc(Al$_2$Cu)$_4$ structure around it effectively lowers the corresponding $E_a$ values, the hopping time can be significantly reduced, ranging from the order of $10^5$ (~31 hours) to less than one second across 200 to 400 °C. Given that the hopping times are generally in the order of seconds or longer, these interstitial Sc hoppings belong to typical rare events. Notably, it is currently infeasible to compute the exact $E_a$ for every possible hopping across all local atomic configurations during the late-growth stages. Therefore, our results focus on a qualitative analysis of the basic kinetic features of these two stages to determine the key factor causing the temperature-dependent formation of the V phase.

Although the first interstitial hopping event of Sc at 200~400°C requires a long incubation time, local lattice distortions induced by temperature or compositional fluctuations may lower the $E_a$ value, facilitating its occurrence. However, in the late-growth stage, the hopping time at low temperatures within 200~300 °C ranges from several thousand to a few hundred seconds, which is insufficient to enable the widespread formation of the V-Sc(Al$_2$Cu)$_4$ structure under typical heat-treatment durations. Moreover, temperatures beyond 300 °C also raise vacancy concentrations in both the matrix and Ω nanoplates. This facilitates Sc migration from the matrix to the ledges on Ω nanoplates terraces and then into the nanoplates, ensuring sufficient Sc for subsequent structural transitions. Therefore, the in-situ formation of the V-Sc(Al$_2$Cu)$_4$ phase within the Ω nanoplates demands high temperatures exceeding 300 °C to accelerate precipitation, with kinetic factors primarily driving this temperature-dependent behavior.

### 3.3 Atomic configurations of sandwich V nanoplates and evolution process

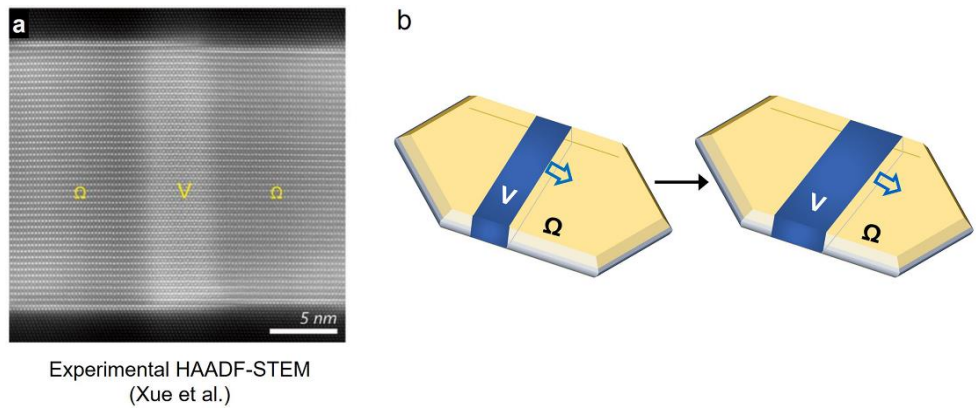

**Figure 7**. a. Experimental HAADF-STEM image, reported by Xue et al. [12], depicting the preferential growth of the V phase within Ω nanoplates. b. Schematic representation of the preferential growth behavior of the V phase within Ω nanoplates.

When the V-Sc(Al$_2$Cu)$_4$ phase forms within Ω nanoplates, a distinct preferential growth behavior can be observed, with the V phase initially forming along the direction perpendicular to the broad terrace of the Ω nanoplates and then expanding laterally to fill them, as depicted in Fig. 7. It is worth noting that the Ω nanoplates have a unique U$_T$|K|θ$_d$-Al$_2$Cu|K|U$_T$ sandwich configuration (see Fig. 1), where the core θ$_d$-Al$_2$Cu part provides dodecahedral interstitial sites, while the outermost shell part (designated as Shell$_Ω$), U$_T$|K layers, with a topologically close-packed (TCP) structure offers only tetrahedral interstitial positions [56,57]. As a transition region between θ$_d$-Al$_2$Cu and the Al matrix, our prior study[58] indicates that the outermost U$_T$|K layers serve as a structural template for Ω nanoplate formation, kinetically inducing the coherent precipitation of θ$_d$-Al$_2$Cu along the habit {111}$_{Al}$ planes. Consequently, due to the lack of dodecahedral interstitial sites, the uptake of Sc into the Ω nanoplates cannot trigger their shell $U_T|K$ layers to transform into a V-Sc(Al$_2$Cu)$_4$/$D2_b$ structure.

In the $\theta_d$-Al$_2$Cu structure, as depicted in Fig. 8a, two distinct types of dodecahedral interstitial positions are evident: Type I (represented by pink-red spheres) and Type II (denoted by black spheres). The $\theta_d$-Al$_2$Cu→V-Sc(Al$_2$Cu)$_4$ transformation requires Sc atoms to migrate into these two types of dodecahedral interstitial sites, respectively. However, within the context of $\theta_d$-Al$_2$Cu in Ω nanoplates, the occupation of Sc atoms may have different schemes due to both the upper and lower sides of $\theta_d$-Al$_2$Cu adjoining $U_T|K$ layers. For example, an Ω nanoplate containing integral multiples of $\theta_d$ unit cells (e.g., 2 $\theta_d$-Al$_2$Cu units, as shown in Fig. 8b) allows for two distinct occupation schemes, designated as integer-type occupation schemes I and II, respectively. The relaxed configurations of these schemes are shown in the lower panel. To determine the stability of these configurations, we introduced an equation defined as:

$$(E_V - E_\Omega)/n_{Sc} = E_{decrease}, \tag{9}$$

where $E_V$ represents the total energy of the model after $\theta_d$-Al$_2$Cu transforms into V-Sc(Al$_2$Cu)$_4$, $E_\Omega$ is the total energy of the model containing an Ω nanoplate, and $n_{Sc}$ denotes the number of Sc atoms incorporated. Based on this definition, $E_{decrease}$ quantifies the energy reduction resulting from the incorporation of Sc atoms into Ω nanoplates. A lower value of $E_{decrease}$ indicates a more stable configuration.

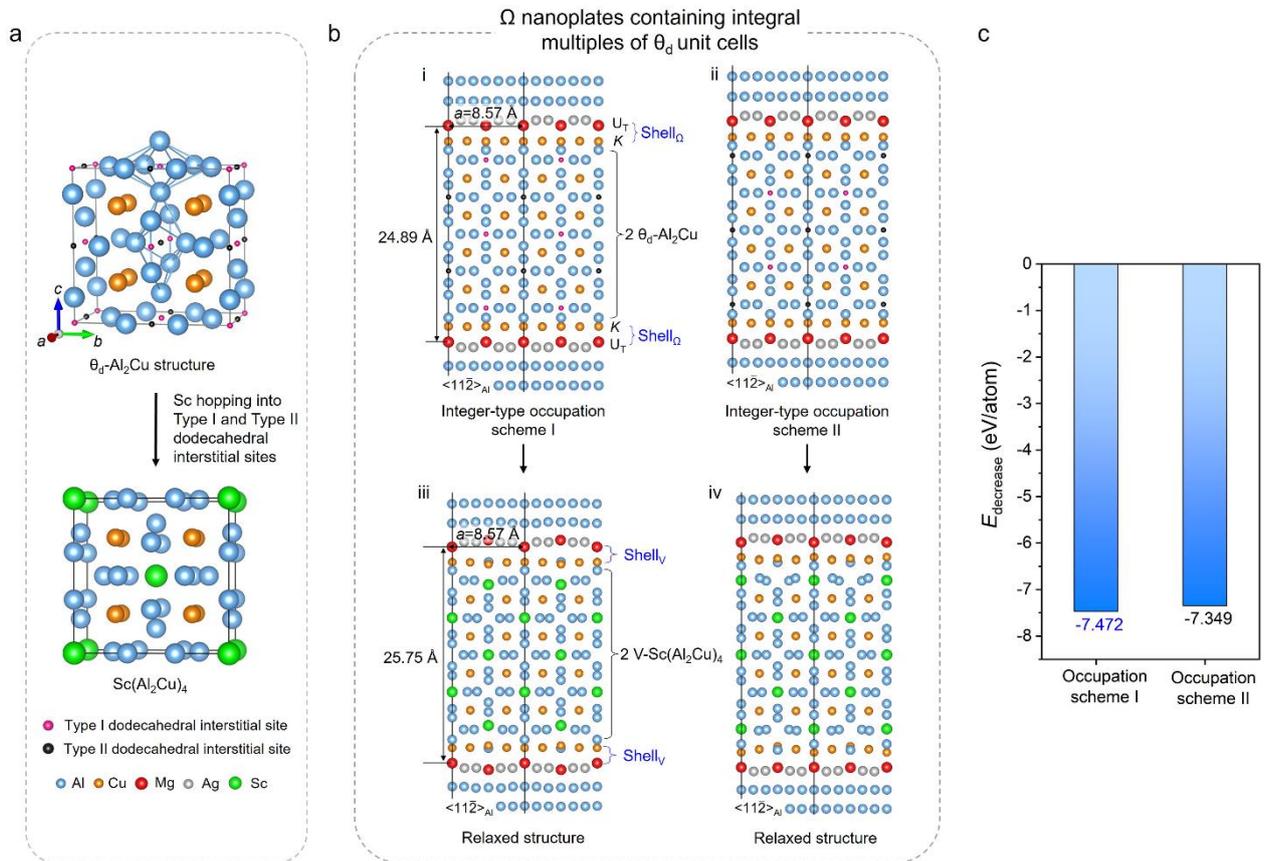

**Figure 8**. a. Schematic illustration of Type I (pink-red spheres) and Type II (black spheres) dodecahedral interstitial

sites in the $\theta_d$-Al$_2$Cu structure. b. Two occupation schemes of Sc atoms within an Ω nanoplate containing 2 $\theta_d$-Al$_2$Cu units in height, with corresponding relaxed structures shown in the lower panel. c. Computation results of $E_{\text{decrease}}$ values for the two integral-type occupation schemes.

The computational results presented in Fig. 8c show that integer-type occupation scheme I yields more negative $E_{\text{decrease}}$ values, indicating that this scheme is more desirable for Sc occupations. Notably, the complete transformation of $\theta_d$-Al$_2$Cu into V-Sc(Al$_2$Cu)$_4$ in the direction perpendicular to the broad terrace of the Ω nanoplates also modifies the original shell parts, U$_T$|K layers, into new configurations, i.e., the newly formed sandwich V nanoplates feature a shell configuration distinct from that of their precursor Ω nanoplates. We designate this novel shell configuration as Shell$_V$, which is composed of $U_T^V$ and $K_T$ lattice nets. A detailed introduction to these lattice nets will be provided in the subsequent section. Furthermore, the structural stability of this sandwich Shell$_V$|V-Sc(Al$_2$Cu)$_4$|Shell$_V$ configuration (Fig. 8b (iii)) at elevated temperatures was confirmed in Fig. S6(a-c) by using the AIMD simulations to equilibrate this structure at 675 K for 10 ps. Therefore, the entire configuration of the newly formed sandwich V nanoplate, containing two multiples of Sc(Al$_2$Cu)$_4$ units, should be as depicted in Fig. 8b(iii).

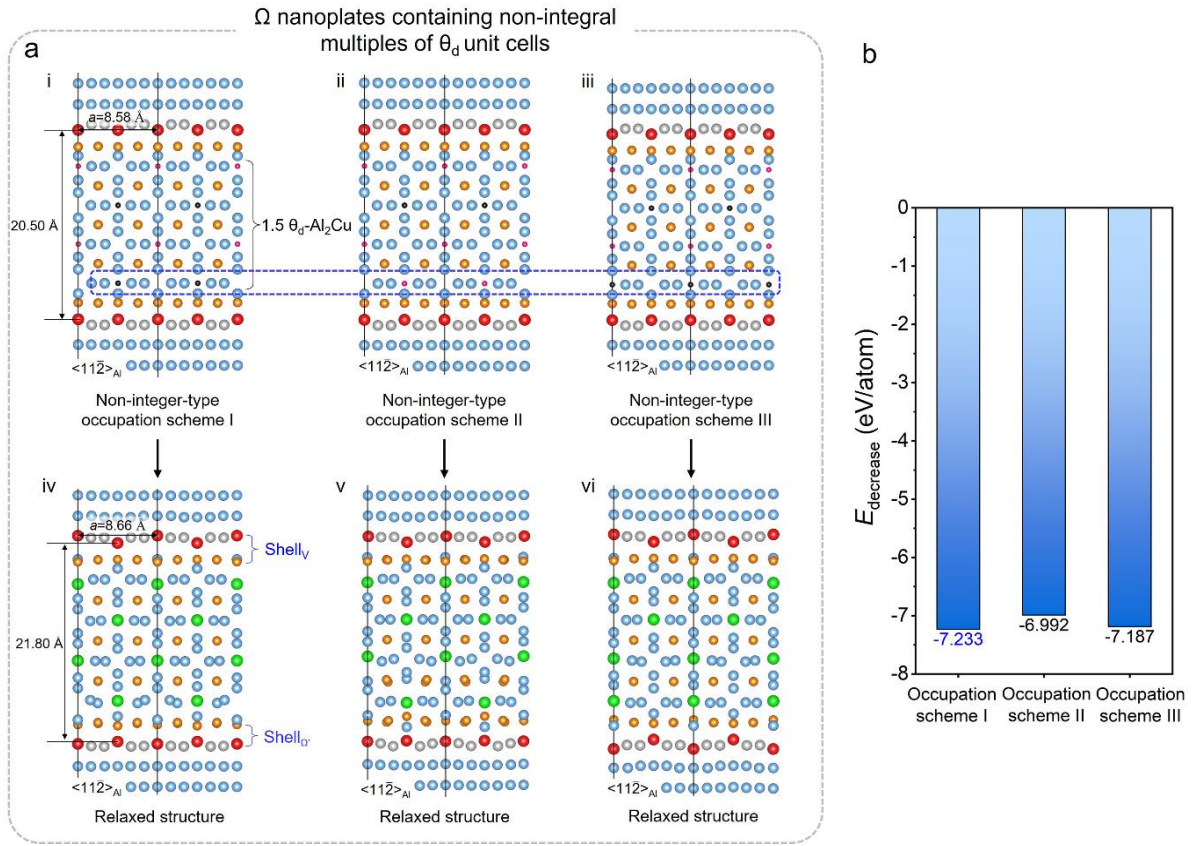

**Figure 9**. a. Three occupation schemes of Sc atoms within an Ω nanoplate containing 1.5 $\theta_d$-Al$_2$Cu units in height, with corresponding relaxed structures shown in the lower panel. c. Computation results of $E_{\text{decrease}}$ values for the three non-integral-type occupation schemes.

For Ω nanoplates with non-integral multiples of $θ_d$ unit cells, taking an Ω nanoplate containing 1.5 $θ_d$ unit cells as an illustration, Fig. 9a presents three distinct occupation schemes, designated as non-integer-type occupation schemes I, II, and III, respectively. The main difference among these schemes is the occupation position of Sc atoms in the region adjacent to the K|$U_T$ layers, as marked by a dark-blue dotted rectangle. According to Eq. (9), Fig. 9b reveals the non-integral-type occupation scheme I has the most negative $E_{decrease}$ values, indicating that the most energetically favorable V nanoplate configuration should be as shown in Fig. 9a (iv). This configuration differs from the structure presented in Fig. 8b(iii), featuring a sandwich V nanoplate with two distinct shell configurations: the upper shell adopts a $Shell_V$ structure, while the lower shell displays a $Shell_{Ω'}$ structure, which is very similar to the original $Shell_Ω$ structure but with slight variations, as detailed in the subsequent section. Using AIMD simulations, the structural stability of this $Shell_V$|V-Sc(Al$_2$Cu)$_4$|$Shell_{Ω'}$ configuration at elevated temperature was also verified in Fig. S6(d-f). Consequently, these results suggest that sandwich V nanoplates with non-integral multiples of Sc(Al$_2$Cu)$_4$ units tend to maintain two distinct shell configurations.

Figure 10 presents simulated HAADF-STEM images of the newly proposed $Shell_V$ and $Shell_{Ω'}$ configurations in sandwich V nanoplates, comparing them with experimental data. In both [001]$_V$ and [010]$_V$ projections, the simulated HAADF-STEM images of these two shell configurations exhibit very similar bright contrast, with no discernible differences, closely aligning with experimental observations. This alignment confirms the reliability of the proposed shell configurations; however, it also exposes the limitations of current HAADF-STEM techniques in distinguishing between these shell structures. Consequently, additional DFT calculations are essential to reveal the underlying structural differences. After the complete transformation of Ω nanoplates into V nanoplates, the thickness of the V nanoplate increases relative to the initial Ω nanoplates. As illustrated in Fig. 8b(iii), the thickness of an Ω nanoplate containing 2 $θ_d$-Al$_2$Cu units increases from 24.89 Å to 25.75 Å, and Fig. 9a(iv) shows a similar increase for another type of Ω nanoplate from 20.50 Å to 21.80 Å, while the $a$ values of both configurations remain nearly unchanged. Given that the formation of Ω nanoplates mainly produces contraction strain in the direction of their thickness [14,59] (i.e., generating a vacancy-type strain field in the upper and lower Al matrix regions), the in-situ formation of the V-Sc(Al$_2$Cu)$_4$ phase within these nanoplates will alleviate this contraction strain, thereby facilitating preferential growth of the V phase perpendicular to the broad terraces of the Ω nanoplates.

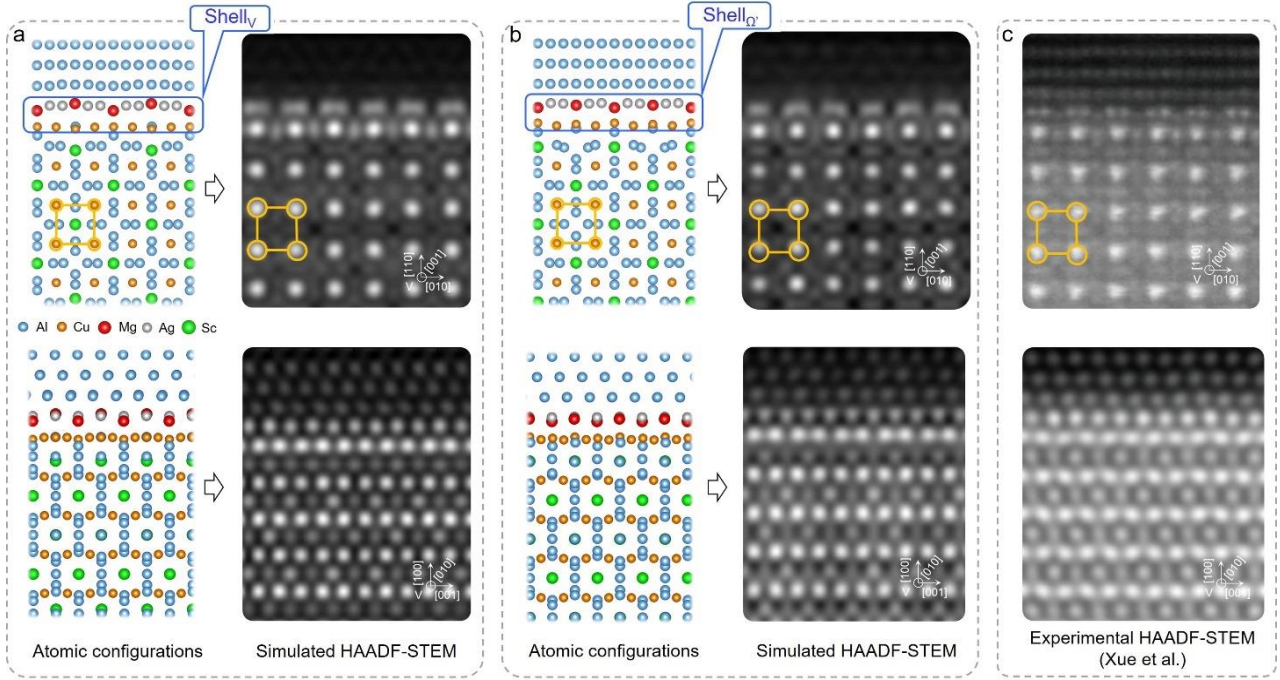

**Figure 10**. a-b. Simulated HAADF-STEM images of the Shell$_V$ and Shell$_{\Omega'}$ configurations for the sandwich V nanoplates. c. Experimental HAADF-STEM images, reported by Xue et al.[12], depicting the atomic structure of the interface between the sandwiched V nanoplates and the Al matrix from the [001]$_V$ and [010]$_V$ directions.

### 3.4 The origin of exceptional thermal stability in sandwich V nanoplates

Compared to the limited thermal stability of Ω nanoplates, which coarsen significantly above 200 °C, the sandwich V nanoplates resulting from the θ$_d$-Al$_2$Cu→V-Sc(Al$_2$Cu)$_4$ transition exhibit extremely excellent heat resistance. Even after long-term heating at 400 °C, the size, number density, and volume fraction of these V nanoplates remain nearly unchanged. Typically, coherent precipitates in alloys tend to coarsen under prolonged high-temperature exposure [4,11,59]. However, no evident coarsening is observed in the sandwich V nanoplates, highlighting their remarkable coarsening resistance. This study elucidates that the unique thermal stability of sandwich V nanoplates arises from alterations in their outermost shell configurations, differing from the original Shell$_\Omega$ structure, which effectively inhibits thickening along the pathway of their precursor Ω nanoplates. The high structural similarity between θ$_d$-Al$_2$Cu and V-Sc(Al$_2$Cu)$_4$ enables initially formed Ω nanoplates to serve as structural templates, remarkably facilitating the coherent precipitation of more complex V nanoplates within the Al matrix through slight local lattice adjustments. Without this templating effect, the direct precipitation of $D2_b$-structured V nanoplates from the parent Al matrix is kinetically unfavorable and inevitably involves severe local lattice reconfigurations. Consequently, the formation of Ω nanoplates is a prerequisite for the subsequent coherent precipitation of V nanoplates.

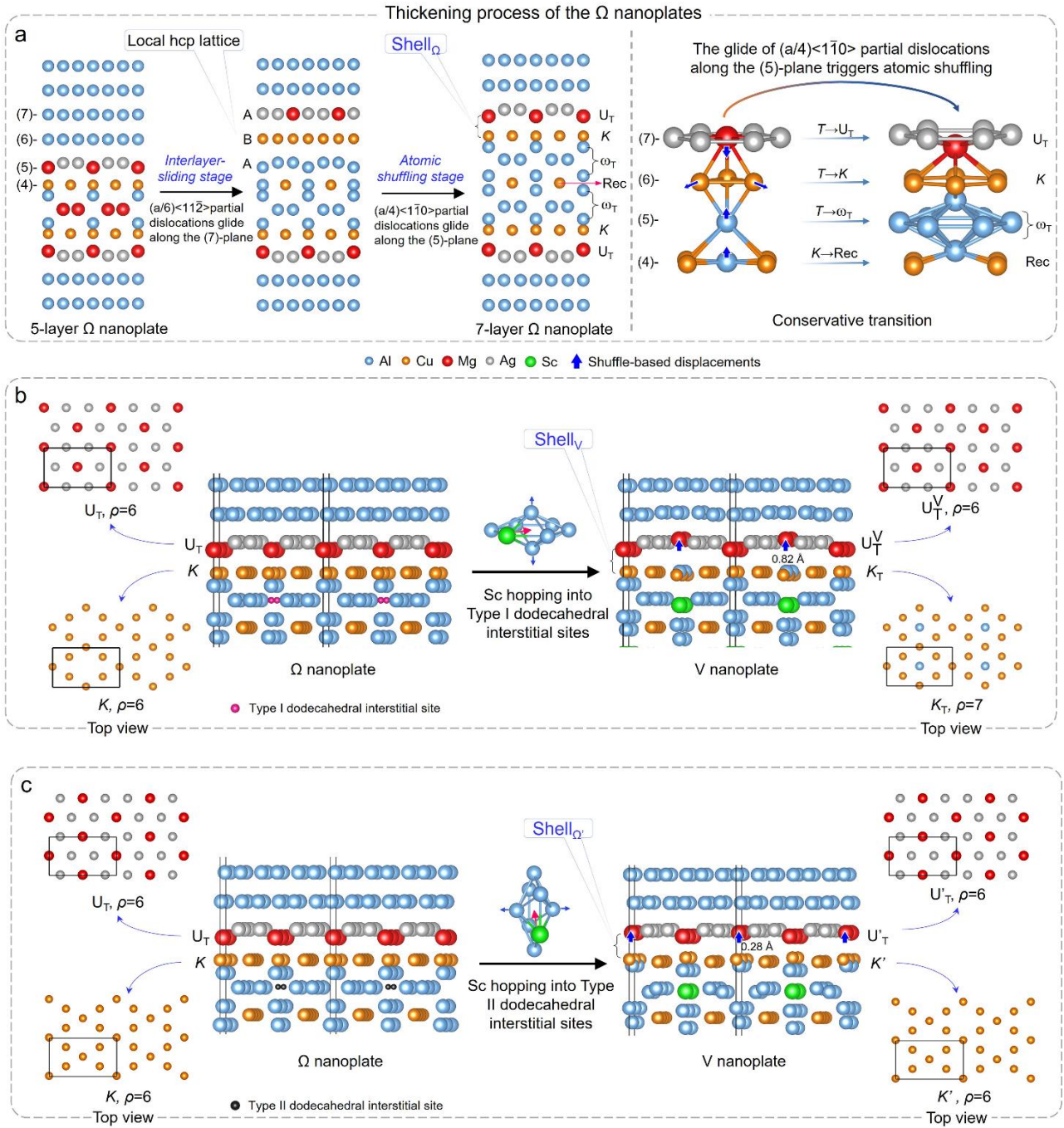

**Figure 11**. a. Schematic of the thickening process of Ω nanoplates driven by interlayer sliding and atomic shuffling stages. The right panel shows a detailed atomic shuffling process across the (4-7)-planes, triggered by the glide of (a/4)<1$\bar{1}$0> partial dislocations along the (5)-plane. b-c. Atomic configurations of the Shell$_V$ and Shell$_{Ω'}$ in the newly formed V nanoplate, with top views of the corresponding lattice nets presented.

According to our recent study [58], the thickening/lengthening of Ω nanoplates is a conservative transformation that does not require the participation of additional atoms. As shown in Fig. 11a, the thickening from a 5-layer to a 7-layer Ω nanoplate occurs via sequential interlayer sliding and atomic shuffling, driven by (a/6)<11$\bar{2}$> and (a/4)<1$\bar{1}$0> partial dislocations, respectively. First, the redistribution of composition from the (4, 5)-planes to the (6, 7)-planes creates a favorable chemical environment that facilitates the nucleation and glide of $(a/6)<11\bar{2}>$ partial dislocations along the (7)-plane, which introduces a local hcp lattice within the (5-7)-planes. Subsequently, the glide of $(a/4)<1\bar{1}0>$ partial dislocations along the (5)-plane triggers atomic shuffling (dark-blue arrows, right panel of Fig. 11a), directly forming the 7-layer Ω nanoplate. During this process, a portion of the atoms in the (4)-plane moves outward to form a $ω_T$ lattice net with the (5)-plane, and the rest form a Rec lattice net (see Fig. A1(b-d) for details on these lattice nets). In this way, $θ_d$-$Al_2Cu$ in Ω nanoplates can add two new layers of material, and the outermost shell $U_T|K$ layers move outward by two $\{111\}_{Al}$ planes simultaneously, without requiring any additional atoms throughout the entire process.

Once Ω nanoplates fully transform into V nanoplates, not only does the core $θ_d$-$Al_2Cu$ change to the V-$Sc(Al_2Cu)_4$ structure, but the original shell $U_T|K$ layers reconfigure, forming $Shell_V$ or $Shell_{Ω'}$ configurations. As the formation of $Shell_V$ structure schematically illustrated in Fig. 11b, the hopping of Sc atoms into Type I dodecahedral interstitial sites (indicated by pink-red spheres) adjacent to the $K$ lattice net pushes the upper and lower Al atoms to move outward (denoted by blue arrows). Notably, the upper Al atoms relocate into the hexagonal opening of the $K$ lattice net, modifying the packing pattern and increasing the atomic density (defined as the number of atoms per unit cell plane) of this lattice net from $ρ=6$ to $ρ=7$. Moreover, these movements also drive half of the Mg atoms within the $U_T$ lattice to move outward by 0.82 Å. Here, we designate these newly formed lattice nets in $Shell_V$ as $U_T^V$ and $K_T$, respectively, with their top views shown in the right panel of Fig. 11b. Additionally, the formation of the $Shell_{Ω'}$ configuration, depicted in Fig. 11c, involves Sc atoms hopping into Type II dodecahedral interstitial sites (indicated by black spheres) adjacent to the $K$ lattice net. Unlike in $Shell_V$, the lattice nets in $Shell_{Ω'}$ do not undergo significant changes; the $K'$ lattice net slightly bends, and half of the Mg atoms in the $U'_T$ lattice net move outward by 0.28 Å (denoted by blue arrows). Due to these subtle differences, this shell is designated $Shell_{Ω'}$, with its lattice nets named $U'_T$ and $K'$, respectively.

Given that the thickening of Ω nanoplates is a conservative transition process, the increase in atomic density $ρ$ of the resultant $K_T$ lattice net in $Shell_V$ configuration would inevitably impede further thickening of these sandwich V nanoplates along the pathway of their precursor Ω nanoplates, causing the addition of the new $θ_d$-$Al_2Cu$ material into the V nanoplates infeasible. Moreover, the formation of the $K_T$ lattice net also alters the crystal symmetry of the original $K$ lattice, leading to the $(a/4)<1\bar{1}0>$-type partial dislocations that cannot occur within its adjacent plane, as the dislocation structure is mainly determined by crystal symmetry [60,61]. Although minor variations in lattice nets in $Shell_{Ω'}$ may not effectively inhibit the thickening of V nanoplates, these nanoplates can still achieve superior coarsening resistance once this side shell evolves into $Shell_V$ through further thickening. Therefore, we suggest that it is through such in situ $\theta_d$-$Al_2Cu$→V-$Sc(Al_2Cu)_4$ transformation to modify the shell configuration that the coarsening resistance of resultant V nanoplates can be significantly enhanced at high temperatures.

### 3.5 Screening of $D2_b$-structured nanoplates in Al alloys

In the Al-Cu-Mg-Ag-Sc alloy, each alloying element is indispensable for the coherent and dense precipitation of Ω and subsequent V nanoplates. Our prior study [58] has shown that the Ω phase can be regarded as a distorted θ-$Al_2Cu$ or a variant of the equilibrium θ-$Al_2Cu$ phase in Al-Cu binary alloys. The introduction of Mg and Ag to Al-Cu alloys does not alter the alloys' thermodynamics but modifies the precipitation sequence, offering a kinetically favorable pathway that induces the previous incoherent θ-$Al_2Cu$ phase to form coherently along $\{111\}_{Al}$ matrix planes. Although adding Mg alone to Al-Cu alloys can induce Ω nanoplate formation[11,62], it occurs at a very low volume fraction. The further addition of Ag dramatically facilitates the nucleation of Ω nanoplates, which is typically attributed to the stronger binding ability of Ag with Mg atoms[63], greatly accelerating the formation of Mg clusters and facilitating Ω nanoplate nucleation. Consequently, both Mg and Ag are crucial for the sandwich Ω nanoplate formation.

Furthermore, adding Sc to the alloy enables the initially formed Ω nanoplates to serve as structural templates, facilitating the subsequent in-situ formation of the V phase within them. In this precipitation, Cu and Sc are the major alloying elements that thermodynamically drive ternary V/τ phases formation, while Mg and Ag are microalloying elements ensuring the dense formation of Ω nanoplates kinetically. It is through the synergistic effect of these alloying elements that exceptional coarsening-resistant V nanoplates can be formed. However, the high cost of Sc hinders the widespread application of this alloy. To address this issue, we utilized DFT-calculated ternary convex hull diagrams from the OQMD database [33] and experimental ternary phase diagrams from the MSI Eureka database to screen more economical alternatives (i.e., major alloying element M) to Sc and to discover new alloy systems (e.g., Al-X-M systems without Cu) also capable of forming coherent, $D2_b$-structured ternary nanoplates.

Building upon the insights gained above, it is evident that the formation of sandwiched, coarsening-resistant V nanoplates is not easy, requiring the simultaneous satisfaction of both thermodynamic and kinetic conditions. First, thermodynamically, a $D2_b$-structured τ-$ScAl_7Cu_5$ phase remains stable at high temperatures (>300°C) and forms a tie-line with Al in the convex hull, which not only ensures its phase stability under prolonged heating but also enables its precipitation from the Al matrix. Second, on the kinetic side, the formation of V nanoplates requires the presence of precursor Ω nanoplates acting as a structural template to induce in situ $\theta_d$-$Al_2Cu$→V-$Sc(Al_2Cu)_4$ transition. This structural templating effect provides a smooth kinetic pathway for the V nanoplates to grow into a coarsening-resistant, sandwiched configuration through slight lattice adjustments. Consequently, the precipitation of other $D2_b$-structured ternary phases should also demand these phases to be stable and form a tie-line with Al in high-temperature convex hulls. Additionally, the precipitation of a precursor $Al_2X$ phase is necessary for structurally templating the formation of these novel phases.

Based on these thermodynamic and kinetic requirements, as illustrated in Fig. 12, we designed a three-step screening strategy to identify effective major alloying elements that can generate novel $D2_b$-structured nanoplates. Initially, from the OQMD database, $D2_b$-structured ternary phases containing Al with a hull distance ($E_{hull}$) less than 0.05 eV/atom at zero temperature were screened. This initial screening identified 152 candidates across the Al-X(Cu, Fe, Cr, Mn, Ni, Ag, Re, Pd, and Co)-M systems, as summarized in Table S2. The $E_{hull}$ is often used as an indicator of material synthesizability [43]. Although the exact boundary between synthesizable and non-synthesizable materials is not clearly defined, metastable phases exceeding 0.05 eV/atom are generally considered unsynthesizable [64]. Given that thermodynamically unstable phases at 0 K may be stabilized by entropy at high temperatures, we set $E_{hull}<0.05$ eV/atom at 0 K to maximize the candidate pool in this initial step.

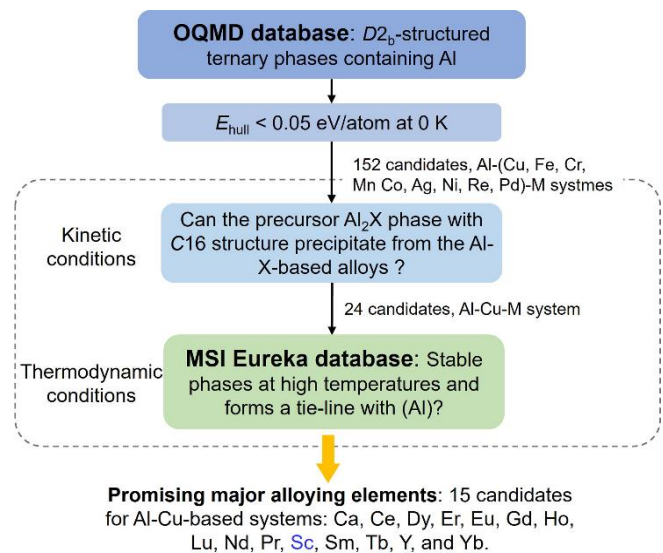

**Figure 12**. Schematic illustrating the procedure to screen effective major alloying elements for producing novel sandwiched nanoplates with a $D2_b$ structure.

In the second screening step, kinetic conditions were examined by evaluating the feasibility of the $Al_2X$ phase precipitation from the Al matrix. For the Al-Cu-based system, our reconstructed Al-Cu convex hull (see Fig. 13a) indicates that the $C16$-structured θ-$Al_2Cu$ phase, although unstable at 0 K with an $E_{hull}$ of 0.023 eV/atom, previous study [45] has demonstrated that it can become stable at temperatures above ~200 °C and forms a tie-line with Al due to stabilization effect of the vibrational entropy. This thermodynamic stability ensures the precipitation of θ-$Al_2Cu$ from Al-Cu-based alloys. However, for other Al-X (Fe, Cr, Mn, Ni, Ag, Re, Pd, and Co)-based systems, their $C16$-structured

Al$_2$X phases are all unstable at 0 K with $E_{hull}$ values largely exceeding 0.05 eV/atom, making their formation thermodynamically infeasible. Thus, only 24 candidates from the Al-Cu-based system passed this step, while others were excluded. Here, we again emphasize that forming the precursor Al$_2$X phase is crucial for the $D2_b$-structured nanoplates. For instance, in the Al-Mn-Ce system, the experimental phase diagram [65] shows that Ce(Al$_2$Mn)$_4$ is a stable phase forming a tie-line with the Al solid solution (designated as (Al)) at 500 °C. However, the corresponding Al$_2$Mn phase cannot precipitate from the Al matrix, as demonstrated in Fig. 13d, which prevents the formation of Ce(Al$_2$Mn)$_4$ phase in a coherent nanoplate structure within the Al matrix; instead, it may appear at grain boundaries in an irregular shape.

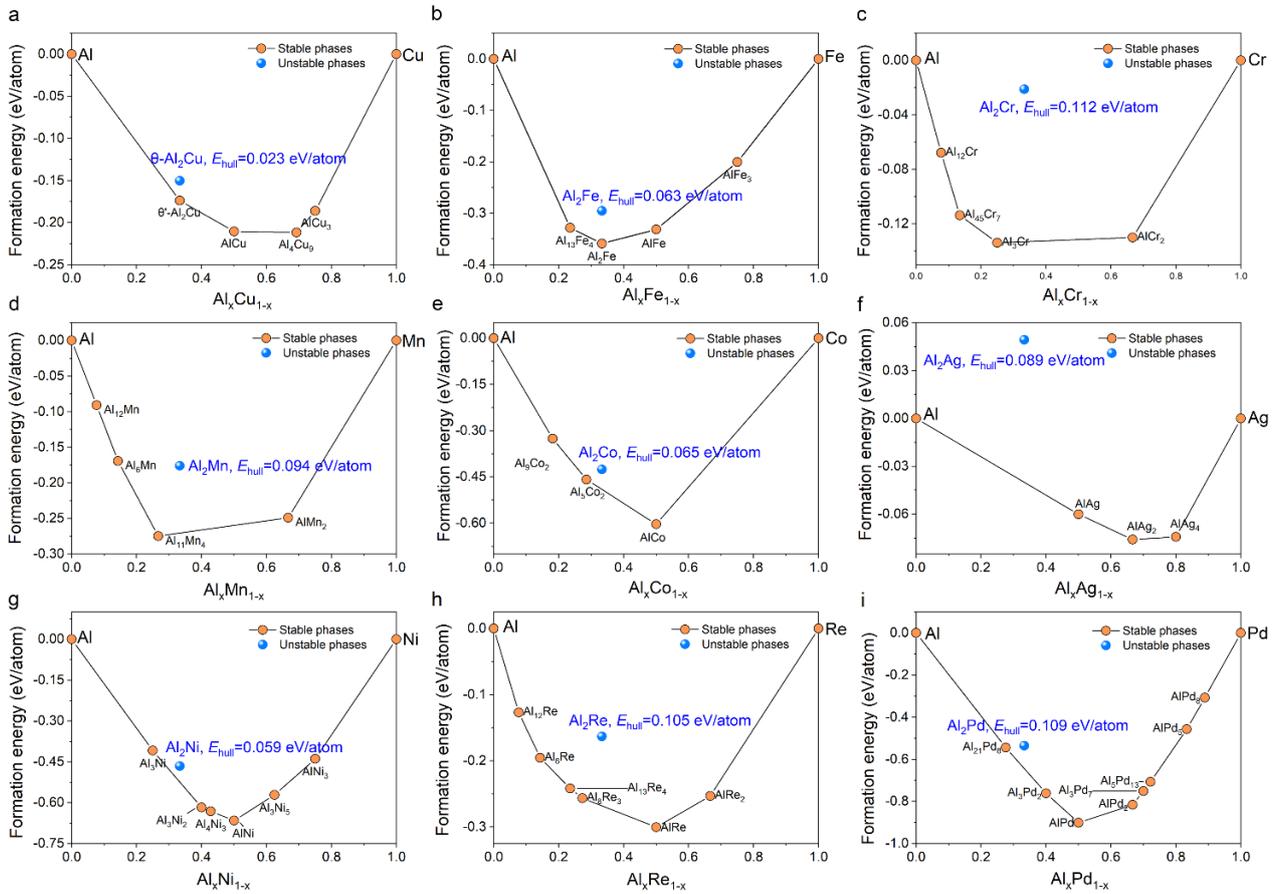

**Figure 13**. The reconstructed zero-temperature convex hull diagrams for Al-X (Cu, Fe, Cr, Mn, Co, Ag, Ni, Re, Pd) systems.

In the final screening, considering that the OQMD database contains only 0 K convex hull diagrams and it is computationally infeasible to rigorously construct high-temperature convex hull diagrams for the remaining 23 Al-Cu-based systems, we thus utilized experimental phase diagrams from the MSI Eureka database to obtain phase stability information at high temperatures (Table 1). During screening, unrecorded elements in Table 1 were excluded as they are mostly radioactive or uncommon, such as U and Tm, and the stable phases that form a tie-line with (Al) were selected.

Through this step, in addition to Sc, we predicted 14 new promising major alloying elements that might substitute Sc, including Ca, Ce, Dy, Er, Eu, Gd, Ho, Lu, Nd, Pr, Sm, Tb, Y, and Yb. At the same time, our results also suggest that the development of novel $D2_b$-structured nanoplates should still focus on Al-Cu-Mg-Ag-based systems. Future research endeavors could explore microalloying elements within these established systems to improve the performance of $D2_b$-structured nanoplates, for instance, by enhancing the volume fraction of nanoplates to further strengthen alloys at high temperatures. However, one should note that current screening offers only preliminary insights, and future experimental investigations should also consider factors like the solid solubility of relevant elements, alloy composition, and thermal processing history.

**Table 1**. Thermodynamic data for $D2_b$-structured ternary $MAl_{8-x}X_{4+x}$ phases at high temperatures (typically 400°C).

| Systems | $MAl_{8-x}X_{4+x}$ phases, M | Stable/Unstable at high temperatures | Can it forms a tie-line with solid solution of Al at high temperatures? (True/False) |
|---|---|---|---|
| Al-Cu-M | Ba | — | — |
| | Ca | Stable | True |
| | Ce | Stable | True |
| | Dy | Stable | True |
| | Er | Stable | True |
| | Eu | Stable | True |
| | Gd | Stable | True |
| | Ho | Stable | True |
| | La | Stable | False |
| | Lu | Stable | True |
| | Nd | Stable | True |
| | Np | — | — |
| | Pm | — | — |
| | Pr | Stable | True |
| | Sc | Stable | True |
| | Sm | Stable | True |
| | Sr | Unstable | False |
| | Tb | Stable | True |
| | Th | — | — |
| | Tm | — | — |
| | U | — | — |
| | Y | Stable | True |
| | Yb | Stable | True |
| | Zr | Stable | False |

**Notes**: "—" in the table indicates that the corresponding experimental ternary phase diagram has not been created. All data in the table are derived from the MSI Eurkea databases.

4. **Summary**

For the exceptional coarsening-resistant V nanoplates in Al-Cu-Mg-Ag-Sc alloys, this study employs comprehensive first-principles calculations to elucidate the thermodynamic properties of the V-Sc(Al$_2$Cu)$_4$ phase, the basic kinetic features during V phase formation within Ω nanoplates, and the mechanisms underlying the high coarsening resistance of V nanoplates. By constructing Al-Cu-Sc convex hull diagrams at 0 K and elevated temperatures, the phase stability of the V-Sc(Al$_2$Cu)$_4$ phase was systematically examined. Unexpectedly, we found that this phase is metastable and thermodynamically tends to evolve into a stable ScAl$_7$Cu$_5$ structure. After carefully accounting for the configurational and vibrational entropy of the ScAl$_7$Cu$_5$ phase, our calculations indicate that the configurational entropy provides a major driving force enabling this phase to form a tie-line with Al in the convex hull and precipitate from the Al matrix. Furthermore, the evaluations of the activation energy for Sc hopping into dodecahedral interstitial positions clarified that kinetics are mainly responsible for the temperature dependence of V phase formation. Through uncovering the entire atomic configurations of sandwich V nanoplates, we found that the formation of V-Sc(Al$_2$Cu)$_4$ within Ω nanoplates changes the $K$ lattice in the shell layer of the Ω nanoplates to a $K_T$ lattice, which hinders the thickening of the V nanoplates along the pathway of the Ω nanoplates, thus contributing to their exceptional coarsening resistance. Finally, using both DFT-calculated convex hull diagrams and experimental phase diagrams, we screened 14 promising substitutes for Sc. This study enhances the understanding of the novel in-situ phase transition stabilization strategy at the atomic level and provides a solid theoretical foundation for developing new creep-resistant Al alloys.

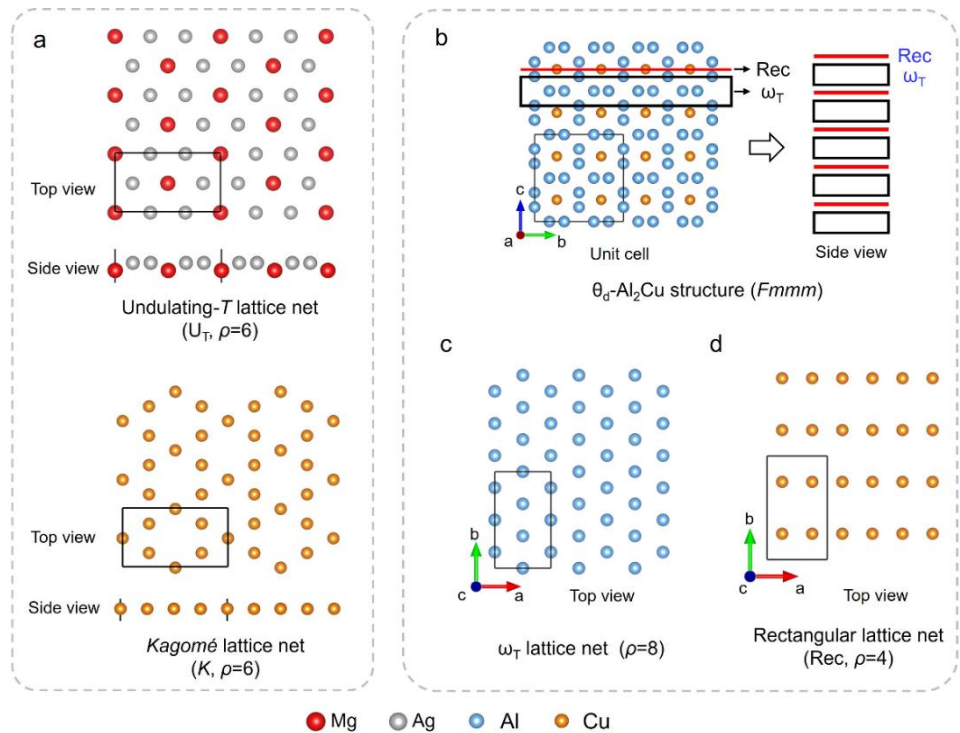

**Figure A1**. a. Atomic packing patterns of the Undulating-T (U$_T$) and Kagomé ($K$) lattice nets. b-d. Schematic depicting the θ$_d$-Al$_2$Cu structure can be stacked by alternating Rectangular (Rec) and ω$_T$ lattice nets.

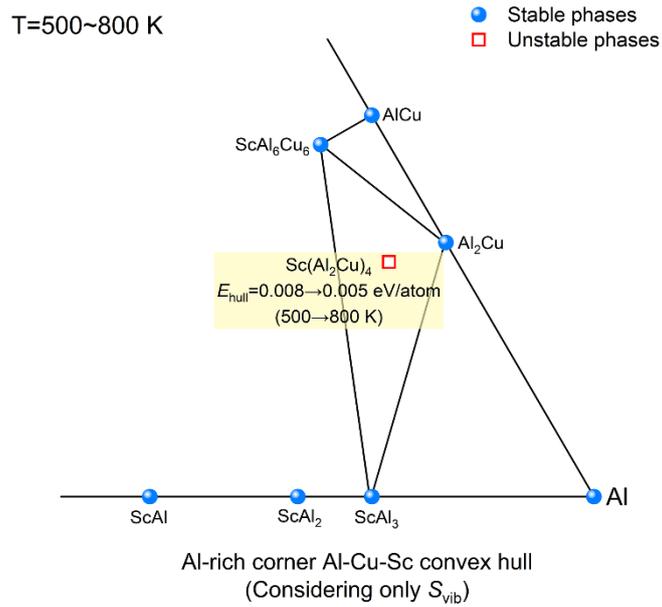

**Figure A2**. The Al-rich corner Al-Cu-Sc convex hull diagram at T= 500~800 K, with $S_{vib}$ considered only in the construction. The detailed $E_{hull}$ values of the unstable Sc(Al$_2$Cu)$_4$ phase are marked in the panel.

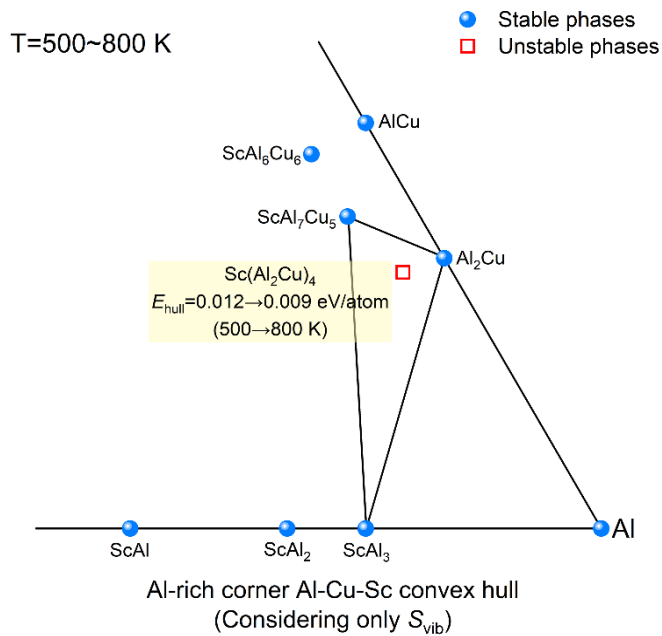

**Figure A3**. The Al-rich corner Al-Cu-Sc convex hull diagram (containing the stable ScAl$_7$Cu$_5$ phase) at T= 500~800 K, with $S_{vib}$ considered only in the construction. The detailed $E_{hull}$ values of the unstable Sc(Al$_2$Cu)$_4$ phase are marked in the panel.

**Acknowledgements**

This research is supported by the National Key Research and Development Program of China (2023YFB3710902), Fundamental Research Funds for the Central Universities of China (N2102011, N2007011, N160208001), and National 111 Project (B20029).

**Author contributions**

G. Qin and J. Bai conceived the original idea and designed the work. J. Bai conducted the simulations with help from X. Pang, Z. Zhao, H. Xue, J. Li, and G. Liu. J. Bai wrote the paper. G. Qin supervised the project and revised the manuscript.

Supplementary Information for
# First-principles Investigation of Exceptional Coarsening-resistant V-Sc(Al$_2$Cu)$_4$ Nanoprecipitates in Al-Cu-Mg-Ag-Sc alloys

Junyuan Bai[1], Hao Xue[1], Jiaming Li[1], Xueyong Pang[1,3], Zhihao Zhao[1], Gang Liu[4], Gaowu Qin[1,2]*,

[1]*Key Laboratory for Anisotropy and Texture of Materials (Ministry of Education), School of Materials Science and Engineering, Northeastern University, Shenyang 110819, China*

[2]*Institute for Strategic Materials and Components, Shenyang University of Chemical Technology, Shenyang 110142, China*

[3]*Research Center for Metal Wires, Northeastern University, Shenyang 110819, China*

4 *State Key Laboratory for Mechanical Behavior of Materials, Xi'an Jiaotong University, Xi'an, China.*

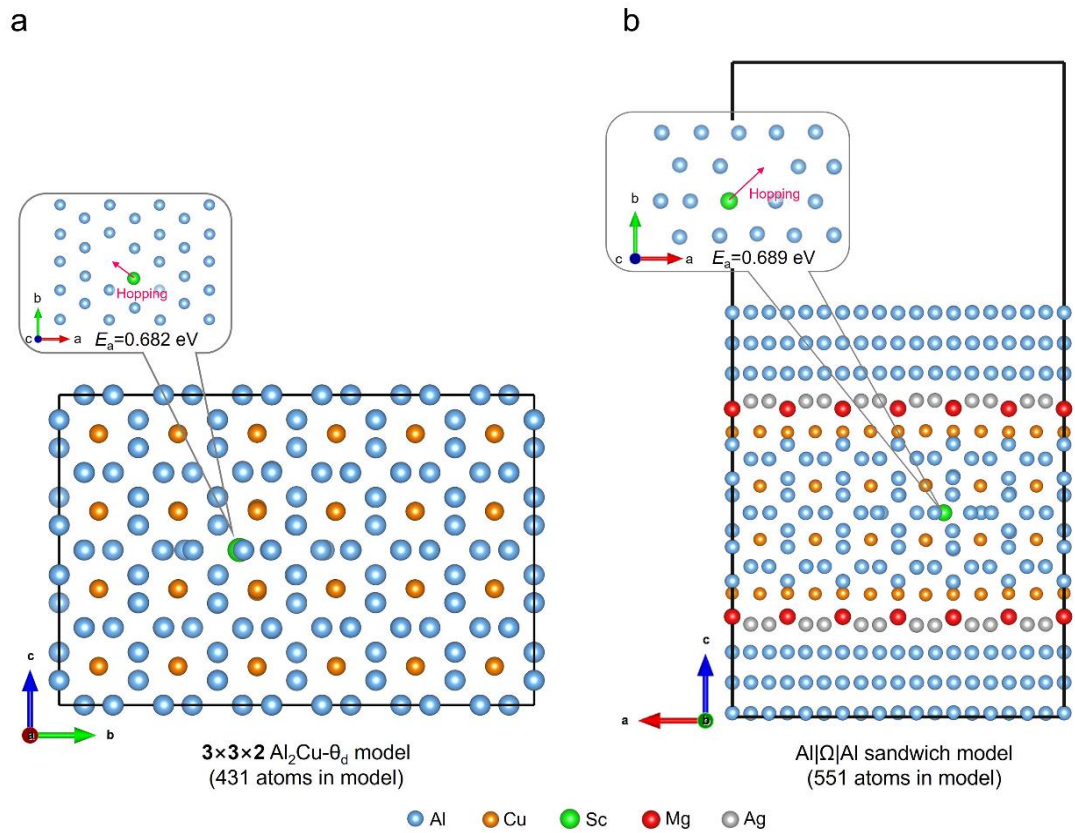

**Figure S1**. Atomic configurations depicting the 3×3×2 Al$_2$Cu-θ$_d$ model containing 431 atoms and the Al|Ω|Al sandwich model containing 551 atoms.

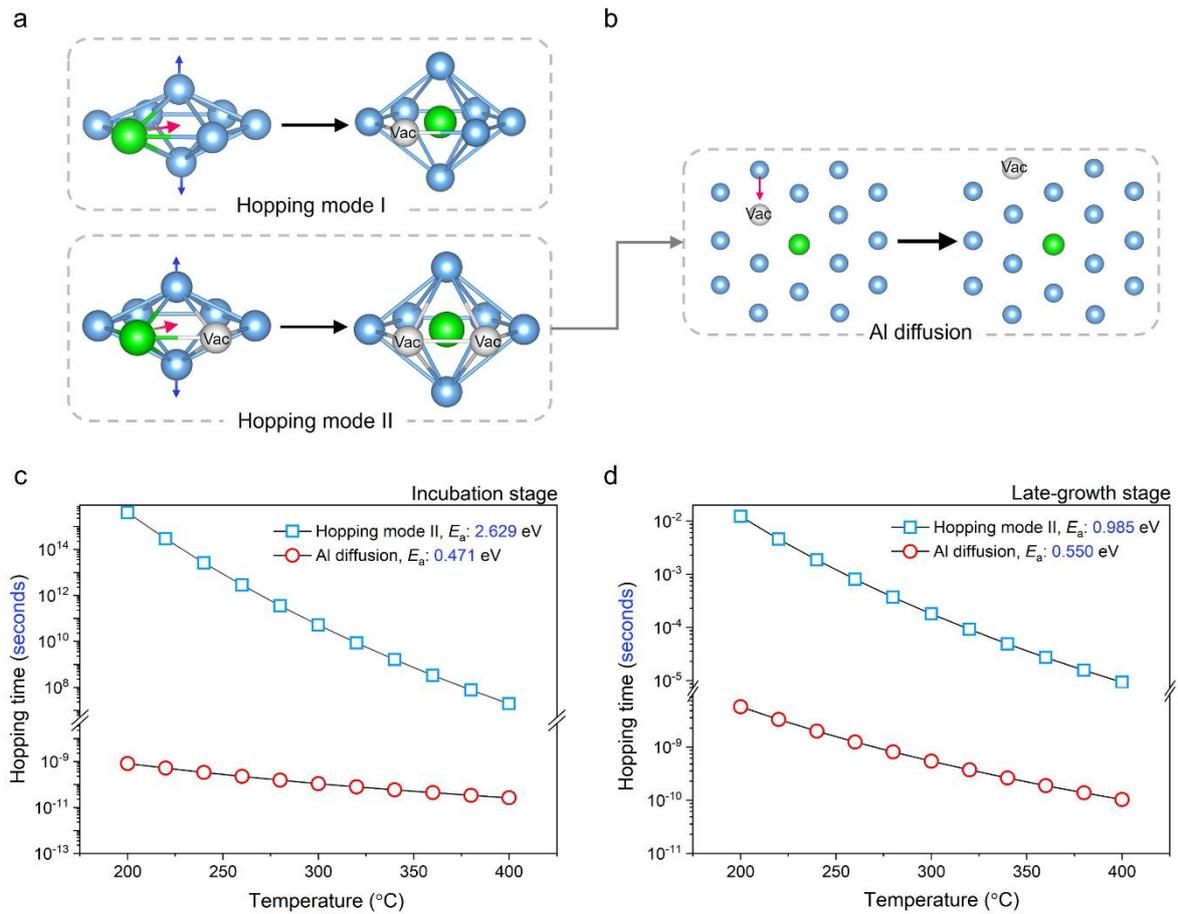

**Figure S2**. **a**. Schematic representation of two possible interstitial hopping modes, designated as mode I and II, for Sc atoms migrating into dodecahedral interstitial positions. **b**. Schematic showing the diffusion of Al atom towards vacancy (Vac) positions adjacent to the Sc atoms. **c**. Temperature-dependent variation in the incubation time for hopping mode II and Al diffusion during the incubation (c) and late-growth (d) stages.

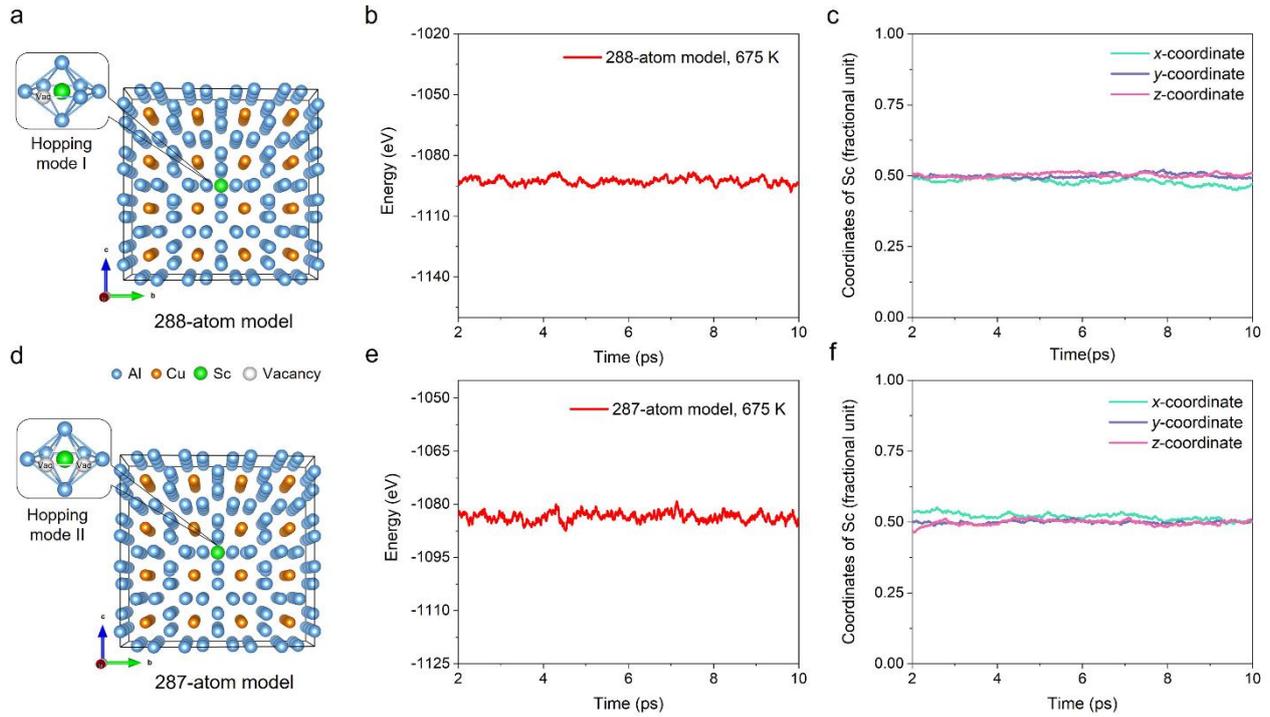

**Figure S3**. (a, d) Atomic models used to examine the structural stability of hopping modes I and II at 675 K. (b, e). Variation in the total energy of the models as a function of time. (c, f) Variation in the coordinates of the Sc atom within the models as a function of time.

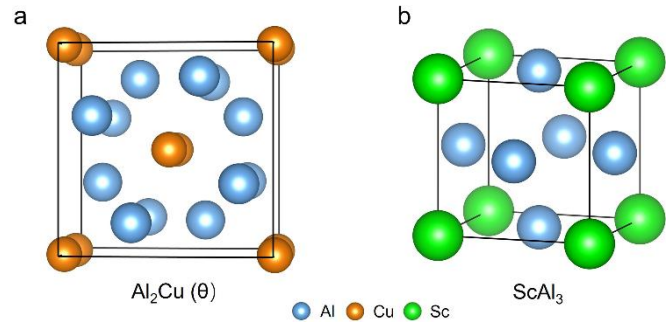

**Figure S4**. The crystal structures of Al$_2$Cu ($\theta$) and ScAl$_3$ phases.

**Table S1**. The calculated formation energies of various antisite defects in the Al$_2$Cu ($\theta$) and ScAl$_3$ phases.

| Phases | Antisite types | Antisite defect formation energies (eV/atom) |
|---|---|---|
| Al$_2$Cu ($\theta$) | Al$^{Cu}$ | 0.188 |
|  | Cu$^{Al}$ | 1.035 |
| ScAl$_3$ | Al$^{Sc}$ | 0.709 |
|  | Sc$^{Al}$ | 2.375 |

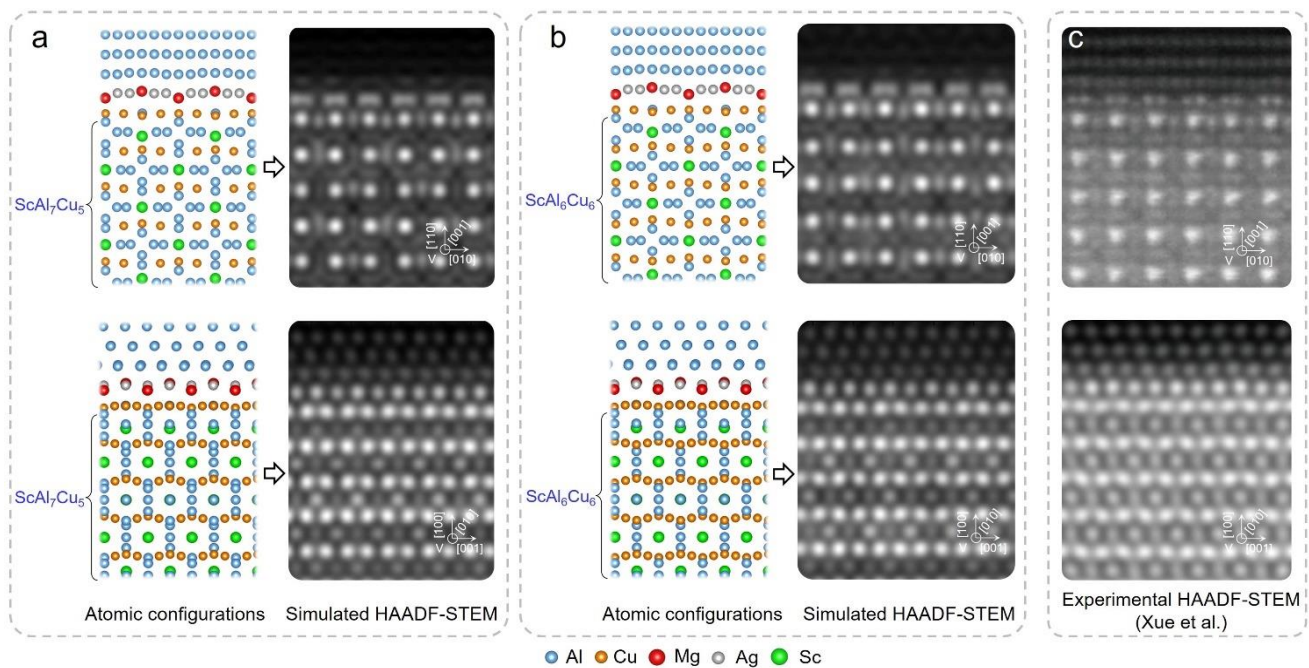

**Figure S5**. Simulated HAADF-STEM images for the sandwich nanoplates containing $ScAl_7Cu_5$ (a) and $ScAl_6Cu_6$ (b) phases. c. Experimental HAADF-STEM images, reported by Xue et al.[1], depicting the atomic structure of the sandwich nanoplates from the $[001]_V$ and $[010]_V$ directions.

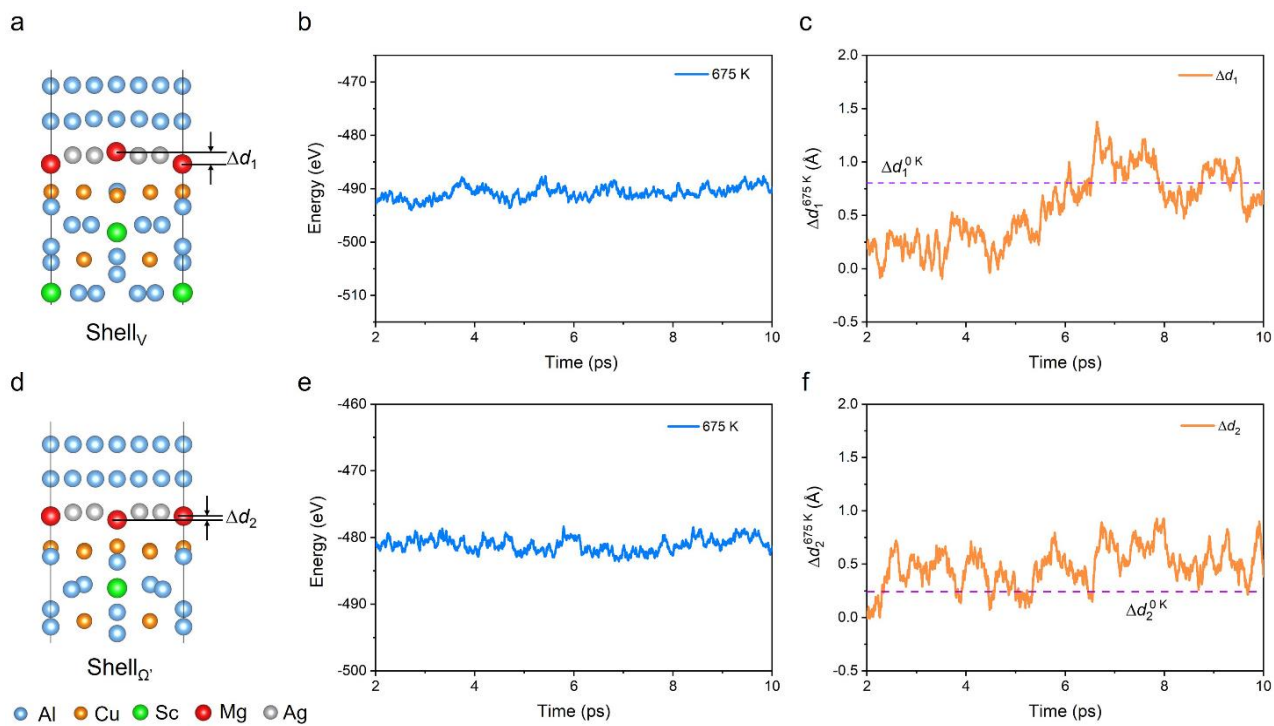

**Figure S6**. (a, d) Atomic models showing the Shell$_V$ and Shell$_{\Omega'}$ configurations. (b, e). Variation in the total energy of the models shown in Figs. 8b(iii) and 9a(iv) at 675 K as a function of time. (c, f) Variation in the $\Delta d_1$ and $\Delta d_2$ as a function of time at 675 K.

**Table S2.** Thermodynamic data for $D2_b$-structured ternary phases at 0 K, obtained from the OQMD [2] databases.

| Al-X-M System | M(Al$_2$X)$_4$ Phases, M | Stable/Unstable ($E_{hull}$, eV/atom) at 0 K | Can it forms a tie-line with Al at 0 K? (True/False) | Al-X-M System | M(Al$_2$X)$_4$ Phases, M | Stable/Unstable($E_{hull}$, eV/atom) at 0 K | Can it forms a tie-line with Al at 0 K? (True/False) |
|---|---|---|---|---|---|---|---|
| Al-Cu-M | Ba | Unstable (0.040) | False | Al-Fe-M | Ca | Stable | False |
| | Ca | Stable | True | | Ce | Stable | False |
| | Ce | Stable | False | | Dy | Stable | False |
| | Dy | Stable | True | | Er | Stable | False |
| | Er | Stable | True | | Eu | Unstable (0.002) | False |
| | Eu | Stable | True | | Gd | Stable | False |
| | Gd | Stable | True | | Hf | Stable | False |
| | Ho | Stable | True | | Ho | Stable | False |
| | La | Stable | False | | La | Stable | False |
| | Lu | Stable | True | | Lu | Stable | False |
| | Nd | Stable | True | | Nb | Unstable (0.022) | False |
| | Np | Unstable (0.045) | False | | Nd | Stable | False |
| | Pm | Stable | True | | Pr | Stable | False |
| | Pr | Stable | True | | Sc | Stable | False |
| | Sc | Unstable (0.016) | False | | Sm | Stable | False |
| | Sm | Stable | True | | Sr | Unstable (0.036) | False |
| | Sr | Unstable (0.012) | False | | Tb | Stable | False |
| | Tb | Stable | True | | Th | Stable | False |
| | Th | Stable | False | | Ti | Unstable (0.013) | False |
| | Tm | Stable | True | | Tm | Stable | False |
| | U | Unstable (0.011) | False | | U | Unstable (0.022) | False |
| | Y | Stable | True | | Y | Stable | False |
| | Yb | Stable | True | | Yb | Stable | False |
| | Zr | Unstable (0.036) | False | | Zr | Stable | False |
| Al-Cr-M | Ca | Unstable (0.049) | False | Al-Mn-M | Ca | Unstable (0.021) | False |
| | Dy | Unstable (0.016) | False | | Ce | Unstable (0.013) | False |
| | Er | Unstable (0.012) | False | | Dy | Stable | False |
| | Gd | Unstable (0.022) | False | | Er | Stable | False |
| | Hf | Unstable (0.021) | False | | Eu | Unstable (0.028) | False |
| | Ho | Unstable (0.013) | False | | Gd | Stable | False |
| | Lu | Unstable (0.010) | False | | Hf | Stable | False |
| | Nd | Unstable (0.041) | False | | Ho | Stable | False |
| | Pr | Unstable (0.047) | False | | La | Unstable (0.011) | False |
| | Sc | Unstable (0.031) | False | | Lu | Stable | False |
| | Sm | Unstable (0.031) | False | | Nd | Unstable (0.001) | False |
| | Tb | Unstable (0.019) | False | | Pr | Unstable (0.006) | False |
| | Th | Unstable (0.014) | False | | Sc | Stable | False |
| | Tm | Unstable (0.011) | False | | Sm | Unstable (0.023) | False |
| | U | Unstable (0.034) | False | | Tb | Stable | False |
| | Y | Unstable (0.016) | False | | Th | Stable | False |
| | Yb | Unstable (0.045) | False | | Tm | Stable | False |
| | Zr | Unstable (0.028) | False | | U | Unstable (0.043) | False |
| Al-Ni-M | Ca | Unstable (0.018) | False | | Yb | Unstable (0.013) | False |
| | Ce | Unstable (0.041) | False | | Y | Stable | False |
| | Dy | Unstable (0.002) | False | | Zr | Unstable (0.008) | False |
| | Er | Unstable (0.001) | False | Al-Re-M | Er | Stable | False |
| | Eu | Unstable (0.027) | False | | Ho | Stable | False |
| | Gd | Unstable (0.005) | False | | Lu | Stable | False |
| | Hf | Unstable (0.028) | False | | Ca | Unstable (0.018) | False |
| | Ho | Unstable (0.003) | False | | Ce | Unstable (0.011) | False |
| | La | Unstable (0.049) | False | | Dy | Unstable (0.024) | False |
| | Lu | Unstable (0.0002) | False | | Er | Unstable (0.031) | False |
| | Nd | Unstable (0.024) | False | | Eu | Stable | False |
| | Pr | Unstable (0.031) | False | | Gd | Unstable (0.015) | False |
| | Sc | Unstable (0.009) | False | | Ho | Unstable (0.026) | False |
| | Sm | Unstable (0.012) | False | Al-Ag-M | Lu | Unstable (0.045) | False |
| | Tb | Unstable (0.004) | False | | Nd | Unstable (0.008) | False |
| | Tm | Unstable (0.00003) | False | | Pr | Unstable (0.006) | False |
| | Y | Unstable (0.003) | False | | Sm | Unstable (0.013) | False |
| | Yb | Unstable (0.008) | False | | Tb | Unstable (0.021) | False |
| | Zr | Unstable (0.028) | False | | Tm | Unstable (0.036) | False |
| | Ca | Unstable (0.0004) | False | | Y | Unstable (0.024) | False |
| | Ce | Unstable (0.015) | False | | Yb | Unstable (0.003) | False |
| | Dy | Unstable (0.002) | False | | Dy | Unstable (0.025) | False |
| | Er | Unstable (0.004) | False | | Er | Unstable (0.021) | False |
| | Eu | Unstable (0.011) | False | | Gd | Unstable (0.035) | False |
| | Gd | Stable | False | | Hf | Unstable (0.024) | False |
| | Ho | Unstable (0.001) | False | | Ho | Unstable (0.023) | False |
| Al-Pd-M | Lu | Unstable (0.007) | False | Al-Co-M | Lu | Unstable (0.015) | False |
| | Nd | Unstable (0.005) | False | | Sc | Unstable (0.026) | False |
| | Pm | Stable | False | | Sm | Unstable (0.041) | False |
| | Pr | Unstable (0.009) | False | | Tb | Unstable (0.028) | False |
| | Sm | Stable | False | | Tm | Unstable (0.018) | False |
| | Tb | Unstable (0.0005) | False | | Y | Unstable (0.041) | False |
| | Tm | Unstable (0.002) | False | | Zr | Unstable (0.028) | False |
| | Y | Stable | False | | | | |
| | Yb | Unstable (0.0001) | False | | | | |

**Supplementary Text 1: Major interstitial hopping mode for Sc atoms driving structural transitions**

Figure S2a shows two possible interstitial hopping modes for Sc atoms toward dodecahedral interstitial positions, designated as mode I and mode II. Specifically, mode I involves Sc atoms hopping into a dodecahedral interstitial site, leaving behind a vacancy at their original site. Mode II represents a scenario where Sc atoms hop into a dodecahedral interstitial site when an adjacent site is already occupied by a vacancy, resulting in two vacancy sites. To investigate the structural stability of these configurations at elevated temperatures once Sc atoms relocate to dodecahedral interstitial positions, we used AIMD simulations, as shown in Fig. S3. The results indicate that the total energy of the system and the coordinates of the Sc atom remain relatively unchanged over 10 ps at 675 K, demonstrating their structural stability.

The detailed hopping times for mode I during the incubation and late-growth stages are presented in Fig. 6. For hopping mode II, our results (see Fig. S2c, d) show that this mode significantly reduces hopping times, ranging from initial seconds at 200 ℃ to microsecond at 400 ℃. However, we found this mode II cannot serve as the major hopping to drive structural transitions. As Fig. S2b shows, if the vacancy migrates outward, mode II will revert to mode I. Hence, through evaluating the hopping time of the Al atoms migrating into this vacancy site in Fig. S2c-d, we found that, in both the incubation and late-growth stages, the corresponding Al hopping times are several orders of magnitude smaller than those of mode II. This means that the vacancy can only stay adjacent to Sc for an extremely short time before migrating to other positions. Therefore, the likelihood of mode II occurring is extremely low, and mode I dominates the entire structural transitions.